\documentclass[iop,apj]{emulateapj}
\usepackage{apjfonts}

\newcommand\teff{{T_{\rm eff}}}
\newcommand\bmi{\hbox{\it B--I}}
\newcommand\lta{\mathrel{\hbox{\raise 0.6 ex \hbox{$<$}\kern
                   -1.8 ex\lower .5 ex\hbox{$\sim$}}}}
\newcommand\gta{\mathrel{\hbox{\raise 0.6 ex \hbox{$>$}\kern
                   -1.7 ex\lower .5 ex\hbox{$\sim$}}}}

\begin{document}

\epsscale{1.1}

\title{Color-Magnitude Diagram Constraints on the Metallicities, Ages,
and Star Formation History of the Stellar Populations in the Carina Dwarf
Spheroidal Galaxy}

\shortauthors{VandenBerg, Stetson, \& Brown}
\shorttitle{Stellar Populations in Carina}

\author{
Don A.~VandenBerg\altaffilmark{1}, 
Peter B.~Stetson\altaffilmark{2}, 
and
Thomas M.~Brown\altaffilmark{3} 
}

\altaffiltext{1}{Department of Physics \& Astronomy, University of Victoria, 
P.O.~Box 1700, STN CSC, Victoria, BC, V8W 2Y2, Canada; vandenbe@uvic.ca}

\altaffiltext{2}{Herzberg Institute of Astrophysics, National Research Council
of Canada, 5071 West Saanich Road, Victoria, BC, V9E 2E7, Canada}

\altaffiltext{3}{Space Telescope Science Institute, 3700 San Martin Drive,
Baltimore, MD 21218, USA; tbrown@stsci.edu}

\submitted{Submitted to The Astrophysical Journal}

\begin{abstract}

Victoria-Regina isochrones for $-0.4 \le$ [$\alpha$/Fe] $\le +0.4$ and a wide
range in [Fe/H], along with complementary zero-age horizontal branch (ZAHB)
loci, have been applied to the color-magnitude diagram (CMD) of Carina.  The
color transformations that we have used have been ``calibrated" so that
isochrones provide excellent fits to the [$(B-V)_0,\,M_V$]-diagrams of M\,3 and
M\,92, when well supported estimates of the globular cluster (GC) reddenings and
metallicities are assumed.  The adopted distance moduli, for both the GCs and
Carina, are based on our ZAHB models, which are able to reproduce the old HB
component (as well as the luminosity of the HB clump) of the dwarf spheroidal
galaxy quite well --- even if it spans a range in [Fe/H] of $\sim 1.5$ dex,
{\it provided} that [$\alpha$/Fe] varies with [Fe/H] in approximately the way
that has been derived spectroscopically.  Ages derived here agree reasonably
well with those found previously for the old and intermediate-age turnoff stars,
as well as for the period of negligible star formation (SF) activity ($\sim
6$--10 Gyr ago).  CMD simulations have been carried out for the faintest turnoff
and subgiant stars.  They indicate a clear preference for SF that lasted several
Gyr instead of a short burst, with some indication that ages decrease with
increasing [Fe/H].  In general, stellar models that assume spectroscopic
metallicities provide satisfactory fits to the observations, including the thin
giant branch of Carina, though higher oxygen abundances than those implied
by the adopted values of [$\alpha$/Fe] would have favorable consequences.

\end{abstract}


\section{Introduction}
\label{sec:intro}

The color-magnitude diagram (CMD) of the Carina dwarf spheroidal (dSph) galaxy
has three prominent features that set it apart from those of similar
systems.  The first, and perhaps most conspicuous, characteristic is the 
remarkable dichotomy of the horizontal branch (HB) into an intermediate-age,
open-cluster-like red clump and an old, metal-poor, globular-cluster-like HB
(see, e.g., Smecker-Hane et al.~1994, Hurley-Keller et al.~1998).  In addition,
deep photometric studies by Bono et al.~(2010), Battaglia et al.~(2012), and
de Boer et al.~(2014b), among others, have established that Carina has a very
thin red-giant branch (RGB), despite the presence of stars that span a range in
[Fe/H] from $\sim -2.9$ to $\sim -1.1$ according to the spectroscopic surveys
carried out by, e.g., Shetrone et al.~(2003), Koch et al.~(2008), Lemasle et
al.~(2012), and Venn et al.~(2012).  Finally, the former investigations
have shown that there is a clear gap in the distribution of stars on the
subgiant branch (SGB) that separate the oldest ($> 10$ Gyr) and intermediate-age
($\lta 7$ Gyr) stellar populations, and possibly on the main sequence (MS) at
$V \sim 22.5, B-V \sim 0.12$ between stars with ages $\lta 2$ Gyr and those
with higher ages.  The evidence that Carina has undergone at least two distinct,
well-separated star formation events over its evolutionary history seems
undeniable.

Because Carina is so distant ($\gta 105$ kpc; e.g., Vivas \& Mateo 2013), 
high-resolution spectra have been obtained for relatively few of its stars,
despite the very considerable effort that has been made by several groups
(including those mentioned above).  As a consequence, the trends of [$m$/Fe]
with [Fe/H] (for individual metals $m$) are not yet as well defined as they
need to be in order for us to achieve a satisfactory understanding of Carina's
star formation history (SFH), and of its chemical evolution over time. 
Nevertheless, the available spectroscopic results suggest that the lowest
metallicity stars in Carina have values of [$\alpha$/Fe] $\approx +0.4$, as
typically found in Milky Way halo stars (e.g., McWilliam 1997).  However, in
contrast with the Galaxy, where such enhancements are characteristic of most
stars with [Fe/H] $\lta -1.0$, the transition to low values of [$\alpha$/Fe]
generally occurs at much lower iron abundances in dSph galaxies (see, e.g.,
Venn et al.~2004, Kirby et al.~2011).  This is usually attributed to low
star formation rates in the latter (e.g., Gilmore \& Wyse 1998), which would
have the consequence that Type Ia supernovae (SN) could begin to contribute to
their chemical evolution before the gas had been enriched to very high [Fe/H]
values by exploding massive stars.

In the case of the oldest Carina stars, the ``knee" in the [$\alpha$/Fe] 
vs.~[Fe/H] relationship appears to occur at [Fe/H] $\sim -2.6$ (de Boer et
al.~2014a), where the plateau at lower metallicities changes into a roughly
linear decrease of [$\alpha$/Fe] to subsolar values at [Fe/H] $\gta -1.4$.
(Where this sequence terminates is not entirely clear due to the small numbers
of observed stars and the considerable scatter in the derived abundances;
also see de Boer et al.~2014b.)  Regardless, because the 
intermediate-age (IA) population includes stars that formed with [Fe/H]
$\sim -1.8$ and [$\alpha$/Fe] $\approx +0.4$, while the oldest population
contains some stars at higher [Fe/H], it is clear that Carina has undergone
inhomogeneous chemical evolution (Venn et al.~2012).
This could plausibly be explained by gas infall (see the discussions
by Lemasle et al.~2012 and de Boer et al.~2014b).  (It is a reasonable
expectation that each major SF event would produce an increase in [$\alpha$/Fe]
--- but not necessarily to same value that is found in the preceding generation
--- which subsequently falls with increasing [Fe/H] as the SF rate declines;
see Gilmore \& Wyse 1991.)

Based on the Mg abundances reported by Lemasle et al.~(2012, also see Venn et
al.~2012), the Mg/Fe number abundance ratios in stars belonging to the IA
population appear to decrease from [Mg/Fe] $\approx +0.4$ at [Fe/H] $\sim -1.8$
to $\approx 0.0$ at [Fe/H] $\sim -1.2$ (but with a large scatter, mostly to
very low values of [Mg/Fe]).  The same studies show that calcium abundances are
considerably more uniform, suggesting no more than a shallow slope in the
[Ca/Fe] vs.~[Fe/H] relation over the same range in iron content.  Because
magnesium has a far bigger impact on stellar models due to its much greater
absolute abundance (see VandenBerg et al.~2012, hereafter V12), Mg is
arguably a better tracer of [$\alpha$/Fe] than Ca.  For this reason and the
above considerations, we have adopted the relations between [$\alpha$/Fe] and
[Fe/H] for the oldest and IA populations of Carina that are shown in Figure 1a.
The abundances for individual stars have been plotted only if they have [Fe/H]
$< -1.8$ or they have [Mg/Fe] values that differ by more than 0.25 dex, at
their [Fe/H] values, from those indicated by the solid curve.  These data
suggest that the dashed locus probably does extend to [Fe/H] $\sim -1.2$ (but
see de Boer et al.~2014b).  Furthermore, the observed large scatter in the
Mg, and hence $\alpha$-element, abundances at [Fe/H] $\gta -1.8$ is a natural
outcome if stars with these iron abundances lie along two, well-separated
[$\alpha$/Fe] versus [Fe/H] relations. 
 
\begin{figure}[t]
\plotone{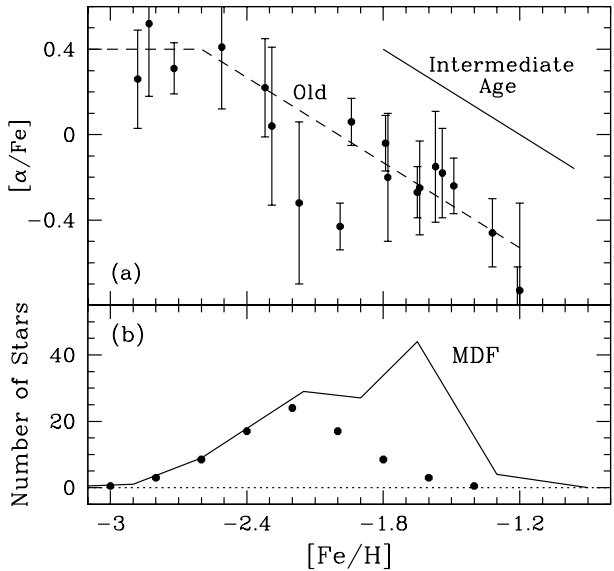}
\caption{{\it Panel (a):} approximate mean [$\alpha$/Fe] versus [Fe/H] relations
 for the oldest and intermediate-age stellar populations of Carina (the dashed
 and solid curves, respectively).  The dashed curve, which reproduces the
 results given by de Boer et al.~(2014a; their Fig.~3),
 clearly provides a good fit to the spectroscopic abundances compiled by Lemasle
 et al.~(2012; see their Table 10), assuming that [$\alpha$/Fe] $=$ [Mg/Fe] (see
 the text).  The solid curve is suggested by a plot of the same data that was
 published by Venn et al.~(2012, their Fig.~20).  Only those stars in the
 Lemasle et al.~sample that have significantly different iron and Mg, and hence
 $\alpha$-element, abundances relative to the values along the solid curve have
 been plotted (the filled circles with attached vertical error bars) because we
 are mainly interested in the ``Old" stars.  {\it Panel (b):} schematic
 reproduction of the Carina MDF that was derived by Starkenburg et al.~(2010).
 The filled circles depict the variation in the number of stars with [Fe/H]
 that has been adopted in this study for the oldest Carina stars (those having
 chemical properties represented by the dashed curve) in the upper panel.}  
\end{figure}

The Carina metallicity distribution function (MDF) that was derived by
Starkenburg et al.~(2010), on the basis of their improved calibration of
Ca\,{\sc ii} triplet absorption line strengths in terms of [Fe/H], is given
in Fig.~1b.  Most of the stars in Carina have
[Fe/H] values between $\sim -2.8$ and $-1.3$, with just a few stars at both
lower and higher metallicities.  Local peaks in the number count distributions
occur at [Fe/H] $\approx -2.2$ and $-1.65$.  Although based on a much smaller
sample, the MDF derived more recently by Lemasle et al.~(2012) agrees quite
well with the Starkenburg et al.~findings.  The main difference between them is
at [Fe/H] $\sim -2.0$, where significantly fewer stars, relative to the peaks on
either side, are present in the Lemasle et al.~sample.  

Insofar as ages are concerned, the narrow RGB of Carina together with the lack
of any obvious metallicity dependence of the colors of red giants at a common
luminosity implies that ages and metallicities are anti-correlated (see, e.g.,
Koch et al.~2006).\footnote{Only recently has it been possible to separate the
most metal-poor giants from those having intermediate metallicities using
measurements of the $c_{U,B,I} = (U-B) - (B-I)$ index (Monelli et al.~2014).
However, as discussed in the latter study, it has proven to be difficult to
explain such observations using current isochrones, perhaps because the assumed
heavy-element mixtures are not representative of the actual stellar abundances.}
Although there have been some attempts to determine the ages of bright giants
using isochrones for close to the spectroscopically derived [Fe/H] values (e.g.,
Lemasle et al.~2012), ``the small offset in color with age $\ldots$ combined
with the sensitivity to the largely unknown elemental abundance mix and He
content limits the usefulness of this exercise" (Koch et al.~2006).  Any
dispersion in the observed abundances of the $\alpha$-elements at the same
[Fe/H] will have important consequences for the effective temperatures (and
presumably the colors) of red giants, at a fixed age (as shown by V12).  Ages
are much more reliably determined for turnoff (TO) and SGB stars since, at low
metallicities, TO luminosity vs.~age relations depend almost entirely on the
absolute abundance of oxygen (and of C and N); the other metals mainly affect
the predicted $\teff$\ scale in a relatively minor way (see V12).
Unfortunately, as noted by Monelli et al.~(2014), very little is known about
the abundances of the CNO elements, and their variations with [Fe/H], in this
(or any other) dSph galaxy.

When this project began (in mid-2014), it was our intention to use the latest
Victoria-Regina isochrones (VandenBerg et al.~2014a; hereafter V14a) for $-0.4
\le$ [$\alpha$/Fe] $\le +0.4$ (and a wide range in [Fe/H]), together with new,
complementary color transformations (Casagrande \& VandenBerg 2014, hereafter
CV14), to study the implications of the very narrow RGB of Carina, and to try
to shed some light on a possible conflict between photometrically inferred and
spectroscopically derived metallicities.  A few years earlier, Bono et
al.~(2010; also see Lianou et al.~2011) had concluded, mainly from the
superposition of Galactic star cluster fiducial sequences onto the Carina CMD
and theoretical simulations of the latter, that the spreads in [Fe/H]
encompassed by both the oldest and the intermediate-age (IA) populations in the
dSph galaxy appear to be considerably less than those derived from spectroscopic
data (also see Fabrizio et al.~2012).

However, there is a third motivation for the present undertaking.  When
interpreting the photometry of complex stellar populations (as in the case of
Carina), it is important to check that the isochrones which are used to
determine their properties (e.g., ages, SFH) faithfully reproduce the CMDs of
simple(r) systems, like globular clusters (GCs).  If they fail to do so, they
cannot be expected to overlie stars in galactic CMDs that have the same
age and chemical abundances, but different
evolutionary states (due to differences in
mass).  In fact, it is a good idea to ``calibrate" stellar models using GCs
and/or open clusters before applying them to complicated systems --- as in the
study of Andromeda's SFH by Brown et al.~(2006) and their recent analyses of the
CMDs for several ultra-faint dwarf (UFD) galaxies (Brown et al.~2014, 2012).
This has generally not been done in studies of Carina, but we have opted to do
so here.  [Note that the ages of most star clusters are not subject to this
concern because they can be derived by matching isochrones to just a small part
of the CMD containing the turnoff luminosity (via, e.g., the procedure
described by VandenBerg et al.~2013; hereafter V13).  In the case of galaxies,
ages are effectively determined on a star-by-star basis, and the isochrones
which are used should provide good fits to photometric data in all of the
evolutionary stages that are studied in order to obtain the most reliable
results.] 

The next section describes the observations that are employed in this
investigation, both of Carina and of the GCs M\,3 and M\,92, which are used to
calibrate the stellar models (see \S~3) that have been applied to the CMD of the
dSph (as reported in \S~4).   Because the most recent Victoria-Regina
isochrones (V14a) are restricted to ages $\gta 5$ Gyr, the main focus of this
project is on the faintest TO stars in Carina, which we have attempted to
explain through the superposition of isochrones onto their observed CMD
locations, as well as numerical simulations.  The latter enable us to probe
the SFH of Carina.  A short summary of our findings is given in \S~5.

%The main purpose of this contribution is to present fits of isochrones
%that allow for the observed mean variations of [$\alpha$/Fe] with [Fe/H] to
%Johnson-Cousins $BVI_C$ photometry of Carina in order to obtain improved
%estimates of the ages of its component stellar populations and to assess how
%well the detailed features of its CMD can be reproduced.  This work should
%extend and improve upon the recent investigation by Bono et al.~(2010), who
%produced a number of synthetic CMDs based on scaled-solar abundance BaSTI models
%for $-2.27 \le$ [Fe/H] $\le +0.06$ (Pietrinferni et al.~2004).  Their study did
%not take into account the observed variations in the abundances of the
%$\alpha$-elements nor the effects of diffusive processes, which have some
%ramifications for age determinations, the morphologies of isochrones in the
%vicinity of the TO, and the luminosity of the HB (due primarily to the larger
%envelope helium abundances that are predicted for the precursor RGB tip stars
%when gravitational settling is not treated).

\section{Photometry of Carina, M\,92, and M\,3}
\label{sec:obs}

The data for Carina discussed here are the same as those that were
the subject of the analysis by Bono et al.~(2010).  They resulted
from the analysis of more than 4,000 CCD images obtained during 27
observing runs between 1992 December and 2008 September.  The
observational indices from each observing run were transformed to
the standard Johnson {\it UBV\/}, Kron-Cousins {RI\/} photometric
magnitude system of Landolt (1992; supplemented by
additional measurements in the {\it UBV\/} filters by Landolt
1973) following the procedures described by, e.g.,
Stetson (2000, 2005).  Then the average results from all observing
runs were used to define a network of local photometric standard
stars in the Carina field, to which observations of all other
stars in the field were referenced.

Exactly the same procedures were employed to produce the
photometric results for our comparision clusters, M\,3 (= NGC\,5272)
and M\,92 (= NGC\,6341).  The results for these two clusters come from
21 observing runs (1984 June--2002 March) for M\,3 and 40 observing
runs (1984 June--2002 May) for M\,92.  

In the final analysis, these results are tied to 324 and 336
Landolt stars in the $B$ and $V$ filters, and 231 Landolt stars in
the $I$ filter.  If we consider the $V$ filter as representative,
the median number of observations per star by Landolt is 16; the
median number of observations per star by us is 69.  We find that
the r.m.s.\ dispersion of the difference between Landolt's and our
average photometric results is 0.012 mag per star in $B$, 0.009
mag in $V$, and 0.011 mag in $I$.  We believe that this represents
an irreducible limit to our ability to transform indices from one
typical set of broad-band filters to another, given the inherent
differences in the detailed spectral energy distributions of stars
of a given perceived color.  The net difference between Landolt's
overall photometric system and our attempt to reproduce that
system should be of order 0.01 mag divided by the square root of a
number of order a few hundred, at least within the range of
temperature, gravity, and chemical abundance that is spanned by
Landolt's (mostly Population I) photometric standards. 

It is true that our photometric study of Carina requires that we
apply transformations derived from field standard stars of mostly
solar metallicity to program stars that are appreciably
metal-poor.  To the extent that different filter sets differently
sample the stellar spectral-energy distributions, this could in
principle introduce a systematic difference between the results of
any one of our filter sets and Landolt's filter set.  Given that
our 27 observing runs employed at least six different combinations
of detector and filter set, this should mitigate any net systematic
difference between the average photometric indices obtained by us
and those that Landolt would have obtained by observing Carina
with his equipment.  In any case, if there does remain a
metallity-dependent difference between what we did measure and what
Landolt would have measured, it should affect Carina, M\,3, and M\,92
similarly, and should be negligible in differential comparisons.
We therefore believe that comparisons between our observations and
our theoretical predictions will be more hampered by imperfect
approximations in the theoretical modelling, by uncertain
transformations between the theoretical and observational indices,
and perhaps by imperfect knowledge of the reddening toward Carina,
than by random or systematic errors in the photometry.

To select our samples of well-observed, probable members of each
system, we first estimated the photometric barycenter of the
target and then plotted the relationship between apparent
magnitude and photometric uncertainty as a function of the radial
distance of the star from the center of its host system.  In
particular, we plotted $V$-band magnitude versus radius for stars
with $0.01 < \sigma(\bmi) < 0.02\,$mag, $0.04 < \sigma(\bmi) <
0.06\,$mag, and $0.08 < \sigma(\bmi) < 0.10\,$mag.  We then chose
an annulus where all of these radial plots were essentially flat. 
Toward smaller radii the errors increase (or, more correctly
stated, the apparent magnitude typical of a given photometric
error becomes brighter) due to crowding; toward larger radii, the
errors increase because those parts of the field typically appear
in fewer CCD images.  As a further limit on the outer radius of
the annulus we plotted observed color versus radius for stars
within a small apparent magnitude range around the cluster
turnoff.  We made sure that stars near the turnoff color dominated
the plot out to our adopted outer limit.  For M\,3 we adopted an
annulus $120 < r < 360\,$arcseconds, while for M\,92 we adopted
$120 < r < 600\,$arcseconds.  For Carina we adopted the entire
disk $r < 600\,$arcsec because we detected no inward increase in
photometric scatter at any radius; apparently, crowding is not a
limiting factor in our data for stars of any brightness within the
magnitude range of interest at any radius all the way in to the
center of the galaxy.

\section{Stellar Evolutionary Models}
\label{sec:models}

Most of the isochrones considered here were derived from the grids of 
evolutionary tracks that have been made publicly available by V14a, using their
interpolation software, and converted to the observed planes via the
transformations produced by CV14.  Both the stellar models and the
color--$\teff$\ relations can be interpolated to arbitrary $\alpha$-element
abundances within the range $-0.4 \le$ [$\alpha$/Fe] $\le +0.4$ at nearly all
of the [Fe/H] values of interest.  (For this study, we have assumed $Y = 0.25$
for the initial mass-fraction abundance of helium, though it is possible to
generate isochrones from the V14a grids for any value of $Y$ between 0.25 and
0.33.) 

We have also made use of a few supplementary grids for [Fe/H] $< -2.4$
that will be the subject of a forthcoming paper on extremely metal-deficient
stars (those having values of [Fe/H] as low as $-4.0$).  Some of these
computations, which allow for variations in [O/Fe] from $+0.4$ to $+1.2$
at fixed [Fe/H] values, assuming [$m$/Fe] $= +0.4$ for the other
$\alpha$-elements, have been employed by Brown et al.~(2014) to interpret their
{\it Hubble Space Telescope} ({\it HST}) photometry of several UFD galaxies.
In addition, we are in the process of generating fully consistent zero-age
horizontal branches (ZAHBs) for all of the aforementioned grids and several of
them appear in the plots that are presented in Section~\ref{subsec:TO}.  (One
of the V14a grids and three ZAHBs have been extended to sufficiently high masses
to permit comparisons of stellar models with the observations of the
intermediate-age TO stars in Carina and with the location of its red HB clump.)

\subsection{Calibration of the Isochrones}
\label{subsec:calib}

As discussed at some length by V13 (also see V14a), there are so many uncertain
factors that play a role in intercomparisons of predicted and observed CMDs
(such as the distance and [Fe/H] scales, heavy-element mixtures, color
transformations, convection theory, atmospheric boundary conditions, the role
of diffusive and extra mixing processes) that ``perfect" agreement between
observations and theory is not a realistic expectation.  It is, in fact,
remarkable that current isochrones fare as well as they do, as the discrepancies
that are found in, for instance, fits of current Victoria-Regina stellar models
to GC photometry (see V14a, their Fig.~14) generally involve no more than
relatively small zero-point and systematic offsets when well-supported estimates
of the cluster properties are assumed.  These offsets are typically at the level
of $\lta 0.025$ mag in the case of $V-I_C$ colors (see CV14) and the very
similar colors which can be derived from the {\it HST} $F606W$ and $F814W$
bandpasses (V13, V14a).

For the most part, isochrones seem to perform nearly as well on the
$(B-V,\,V)$-diagram as on the $(V-I_C,\,V)$-plane (see the examples given by
CV14).  However, as discovered in the course of this study, $B-V$ colors appear
to become increasingly problematic at the lowest [Fe/H] values.  Whereas the
isochrones provide rather good fits to our $BV$ data for M\,3 (comparable to
those presented for M\,5 by CV14) when reasonable assumptions are made
concerning its basic properties (i.e., distance, reddening, and metallicity),
the agreement is noticeably less satisfactory in the case of M\,92.  While the
discrepancies are relatively minor above the turnoff, the models (for [Fe/H] 
$= -2.40$; Kraft \& Ivans 2008; Carretta et al.~2009a) predict an appreciably
steeper MS slope than the observed one.  It seems doubtful that this is due to
problems with the model $\teff$\ scale because the isochrones do not have
similar difficulties when confronting M\,92 data derived from longer wavelength
filters (e.g., $V-I_C$, or the {\it HST} equivalent; see V13, V14a).  It is
even less likely that this difficulty can be attributed to observational errors
because our CMD for M\,92 is based on a very large number of independently
calibrated data sets (see \S~\ref{sec:obs}).  
 
However, a resolution of this issue is not needed for this project.  Since the
$BV$ observations of Carina, M\,3, and M\,92 have all been calibrated to the 
same photometric system, we can force the stellar models to be on very close to
the same system if we simply correct the colors along the isochrones so that
they reproduce the CMDs of the two GCs.  Even though this does tie our analysis
to particular distance and [Fe/H] scales for globular clusters, the resultant
models should yield a more accurate interpretation of the Carina photometry
than those without such adjustments.  One must just be careful to carry out this
calibration of the synthetic colors in such a way that the latter vary smoothly
with [Fe/H] (as well as with $\teff$\ and $\log\,g$) in order that they can be
applied to isochrones for the ranges in age and metallicity found in Carina.

We believe that the best way to achieve this is to set up tables of
$\delta(B-V)$ values to complement the tabulated transformations to $B-V$ given
by CV14.  As these results will be used primarily in the fitting of isochrones
to the faintest TO of Carina, the color correction tables can be restricted to
[Fe/H] $= -3.0(0.5)-1.5$, and at each [Fe/H] value, to $\log\,g = 2.5(0.5)5.0$
and $\teff = 5000(250)6750$~K, where the numbers in parentheses give the
increments to the three variables.  The color--$\teff$ relations provided by
CV14 consider exactly the same values of [Fe/H], $\log\,g$, and $\teff$\ among
their coverage of a greatly expanded parameter space (and $\sim 40$ different
color indices).  They also provide tables for [$\alpha$/Fe] $= -0.4(0.4)+0.4$,
but the majority of the Galactic GCs appear to have [$\alpha$/Fe] $\approx +0.4$
(see, e.g., Carretta et al.~2009b); consequently, the $\delta(B-V)$ adjustments
that are derived from a consideration of the M\,3 and M\,92 CMDs implicitly
assume this value of [$\alpha$/Fe].  (In fact, we have not attempted to
determine how $\delta(B-V)$ depends on the abundances of the $\alpha$-elements.
As shown below, these corrections are, in any case, quite small for the
post-turnoff portions of Victoria-Regina isochrones, and since they are used in
conjunction with the CV14 transformations, which do predict the effects on $B-V$
of differences in [$\alpha$/Fe], this neglect should not be a concern.  In
other words, we have assumed that, at a fixed value of [Fe/H], the {\it
dependence} of $\delta(B-V)$ on $\log\,g$ and $\teff$\ is not a function of
[$\alpha$/Fe].) 

M\,3 and M\,92 are particularly suitable GCs for this exercise because their
[Fe/H] values are close to the grid values of $-1.5$ and $-2.5$, respectively.
Hence, whatever adjustments in the $B-V$ color are needed to obtain good fits
of isochrones to the CMDs of these systems will be very similar to the values
that should appear in the interpolation tables for these metallicities.
Because $B-V$ is much more dependent on $\teff$\ than on $\log\,g$ (especially
for MS, SGB, and lower RGB stars; see, e.g., VandenBerg \& Clem 2003, their
Fig.~3), the temperature shift, at a fixed gravity, between the MS portions of
isochrones for [Fe/H] $= -2.0$ and $-2.5$, as a fraction of that predicted for
[Fe/H] $= -1.5$ and $-2.5$ isochrones of the same age, should be a reasonable
approximation to the corresponding variation in $\delta(B-V)$.  Based on our
stellar models, this fraction is close to 0.30, which we have adopted in the
creation of a table of color corrections for [Fe/H] $= -2.0$ (and for $-3.0$,
using a similar rationale). Indeed, with relatively few iterations, it was
%This sounds complicated and uncertain, but with
%relatively few iterations,} it is possible to 
quite easy to produce a table of $\delta(B-V)$
values that are smooth functions of [Fe/H], $\log\,g$, and $\teff$ and which
enable isochrones to satisfy the M\,3 and M\,92 contraints very well.
Importantly, these corrections are small for the age-sensitive parts of
isochrones --- easily within the uncertainties associated with the stellar
models, color--$\teff$ relations, and photometric zero-points.

\begin{figure}[t]
\plotone{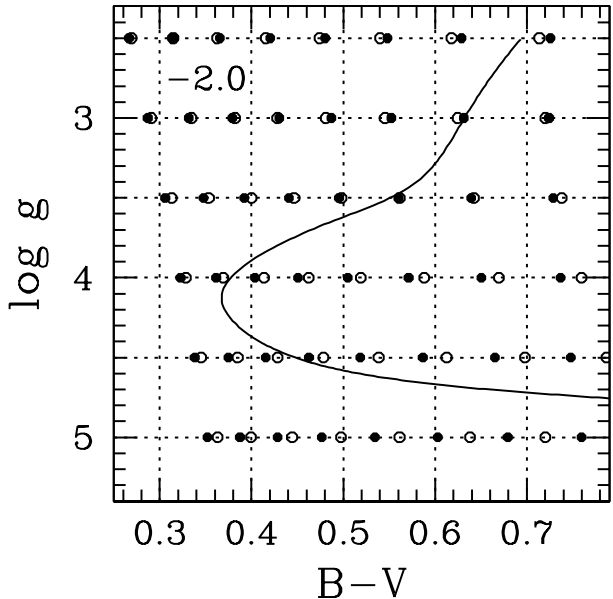}
\caption{Plot of the $(B-V)$--$\teff$--$\log\,g$\ relations for [Fe/H] $= -2.0$,
$\log\,g = 2.5(0.5)5.0$ and, at constant gravity, $\teff = 6750(-250)5000$~K, in  
the direction from left to right.  The filled and open circles give, in turn,
the transformations reported by CV14 and those derived here from fits of
isochrones to the M\,3 and M\,92 CMDs, and other considerations (see the text).
The $\delta(B-V)$ values referred to in the text correspond to the differences
between the open and filled circles at common values of $\log\,g$\ and $\teff$.
The solid curve represents a 12 Gyr isochrone for [Fe/H] $= -2.0$,
[$\alpha$/Fe] $= +0.4$, and $Y=0.25$ (from V14a).}
\end{figure}

This can be seen in Figure 2, which compares the transformations to $B-V$ for
[Fe/H] $= -2.0$ (and [$\alpha$/Fe] $= +0.4$) from CV14 (filled circles)
with those obtained if the latter are corrected by our adopted values of
$\delta(B-V)$ (open circles).  A 12 Gyr isochrone for the same chemical
abundances has also been plotted to permit visual estimates of the variations
in the color adjustments that have been applied as a function of evolutionary
state.  Above the turnoff, they amount to $< 0.02$ mag, rising to $\sim 0.04$
mag only at $\log\,g \gta 4.7$.  Lines of constant $\teff$\ run in a nearly
vertical direction; e.g., the bluest points at the grid values of $\log\,g$
from 2.5 to 5.0 give the predicted colors at $\teff = 6750$~K, whereas the
reddest ones indicate the colors for the coolest $\teff$\ considered (5000~K).
At a fixed gravity, the points at intermediate $B-V$ values illustrate how this
color index varies when the temperature decreases in successive steps of 250~K,
in the direction from left to right.

\begin{figure}[t]
\plotone{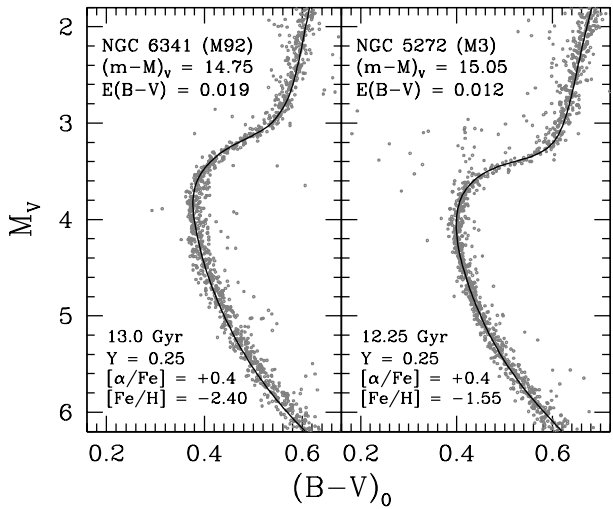}
\caption{Fits of isochrones for the indicated ages and chemical abundances
to our CMDs for M\,92 and M\,3 on the assumption of the apparent distance
moduli (based on ZAHB models) and reddenings (from Schlafly \& Finkbeiner 2011)
that are specified in the upper left-hand corners of the respective panels.  The
predicted $B-V$ colors were derived from the CV14 transformations, but with
small adjustments to the latter that were obtained by interpolation in tables
of $\delta(B-V)$ values for [Fe/H] $= -3.0, -2.5, -2.0$, and $-1.5$ and ranges
of $\log\,g$ and $\teff$ relevant to the faintest TO stars in Carina.  (The
differences between the adopted and CV14 color--$\teff$ relations for stars
having [Fe/H] $= -2.0$ are illustrated in the previous figure.)} 
\end{figure}

Figure 3 demonstrates that our calibration of the CV14 transformations does,
indeed, lead to excellent fits of isochrones to the CMDs of M\,3
and M\,92 (by design, of course), when the reddenings given by Schlafly \&
Finkbeiner (2011) and apparent distance moduli based on ZAHB models for the
indicated chemical abundances are assumed.  (Distances derived in this way are
in very good agreement with those based on the RR Lyrae standard candle; see
V13 and VandenBerg et al.~2014b, hereafter V14b.)  It is worth pointing out that
the same ages would be (and are) found using the CV14 color--$\teff$ relations
because the predicted TO luminosities have not been altered by the adopted
$\delta(B-V)$ offsets.  The main effect of the latter, besides removing small
discrepancies along the lower RGB, is to force the predicted and observed TO
colors (and lower main sequences) to agree.  This concurrence is required, in
fact, if one is to obtain the best estimate of the
turnoff age of a GC from fits of isochrones to just the
small region of its CMD in the vicinity of the TO (see the discussion of this
issue by V13, their \S~3.1).\footnote{The main reason why V13 obtained ages for
M\,3 and M\,92 that are 0.5 Gyr younger than those derived here --- despite
their assumption of slightly smaller distance moduli by 0.03 mag, which would
normally imply {\it increased} ages by $\approx 0.3$ Gyr --- is that the model
grids used in their study assumed higher oxygen abundances by 0.24 dex (in an
absolute sense), due to differences in the respective solar heavy-element
mixtures and the adopted values of [O/Fe].  The higher the absolute oxygen
abundance, the lower the age at a given TO luminosity (see V12).  This example
serves to illustrate the importance of taking chemical composition and distance
modulus differences into account when comparing age determinations for the same
star cluster that are reported in different papers.}  In any case, the colors
along Victoria-Regina isochrones {\it should} be adjusted in the way that we
have described {\it if} the ZAHB-based distance scale is accurate and the actual
cluster [Fe/H] values are close to the values that we have adopted (see Fig.~3).

\begin{figure}[t]
\plotone{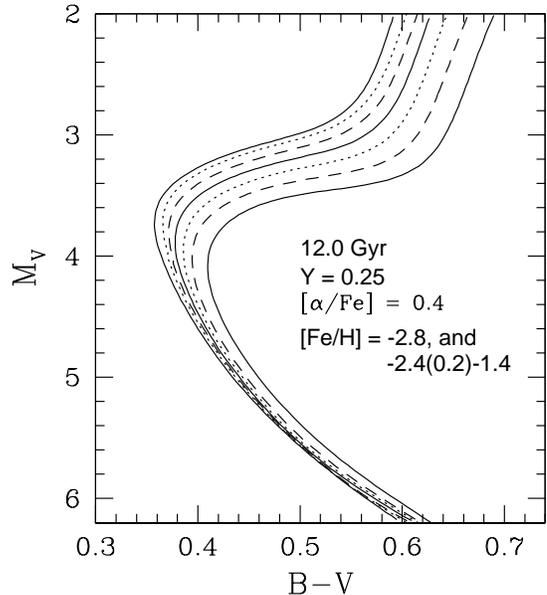}
\caption{Plot of isochrones for the indicated age and chemical abundances if 
the modified CV14 transformations described in this paper are used to transpose
them to the $(B-V,\,M_V)$-diagram.}
\end{figure}

To further illustrate the performance of the calibrated transformations to
$B-V$, we have plotted, in Figure 4, isochrones for seven [Fe/H] values (as
indicated), each assuming $Y = 0.25$, [$\alpha$/Fe] $= +0.4$, and an age of
12 Gyr.  The differences between them,
%(Except for the most metal-deficient one, they have been generated
%from the models provided by V14a, using their interpolation software.  The
%isochrone for [Fe/H] $= -2.8$ has been produced from new grids of tracks, to
%be presented in a forthcoming paper, that will extend the currently available
%Victoria-Regina computations to [Fe/H] $= -4.0$.)  
at a fixed value of $M_V$, vary smoothly and monotonically, and there are no
obvious irregularities as a function of absolute magnitude or color.  The main
effect of the $\delta(B-V)$ values that we have generated is to displace the
lower-MS portions of the isochrones to redder colors, especially at [Fe/H]
$\lta -1.8$, by amounts that vary directly with $M_V$.  As a result, the
isochrones display a considerably reduced sensitivity to [Fe/H] over the range
$4 \lta M_V \lta 6$, as required by the photometry of M\,3 and M\,92 (if our
assumed properties for them are accurate), compared with their behavior when
the CV14 transformations are used without any adjustments. 

Figs.~2, 3, and 4 thus give us considerable confidence that the isochrones used
in this investigation to model the oldest MS, SGB, and lower RGB stars in
Carina are about as well constrained to empirical data as we are able to make
them.  In the case of giant stars (those with $\log\,g \lta 2.5$), the CV14
transformations to $B-V$ appear to require no more than a $-0.01$ mag
adjustment in order for the same isochrones that have been plotted in Fig.~3 to
reproduce the entire RGBs of M\,3 and M\,92 (see \S~\ref{subsec:RGB}, which
presents an analysis of the giant-branch populations of the dSph).  Finally, we
note that the interpolated $\delta(B-V)$ values for $\teff = 6750$~K are used
when the predicted temperatures exceed this limit (as in isochrones that are 
relevant to IA stars), and the tabulated color adjustments for [Fe/H] $= -1.50$
are adopted for slightly more metal-rich stars.  The latter is a reasonable
assumption given that the CV14 transformations appear to provide comparable
fits of isochrones to the CMD of M\,5 (see CV14), which has [Fe/H\ $\sim -1.3$
(e.g., Carretta et al.~2009a), and of the $\sim 0.2$ dex more metal-deficient
cluster M\,3, as we found during the course of this work.  Moreover, there
appear to be very few stars in the faintest TO of Carina with [Fe/H] $> -1.5$
(see Fig.~1a); consequently, they will have no more than a very minor influence
on our findings. 

\section{Interpretation of the Carina CMD}
\label{sec:cmd}

To try to constrain our understanding of Carina, we will first intercompare its
CMD with that of M\,92.  In addition to highlighting some of the similarities
and differences between these systems, this will provide a useful check of the
reliability of the adopted reddenings and distance moduli in a relative sense.
This will be followed by fits of isochrones that utilize the empirically
calibrated color transformations just described to the CMD of the galaxy, to
explore the implications of different assumptions concerning the ages and
chemical compositions of its stellar populations.  Fully consistent ZAHB loci
will also be considered.  Whereas the analyses presented in the first subsection
will reveal some of the possible interpretations of Carina's CMD, with
particular emphasis on the oldest stars, the next one will endeavor to assess
the viability of these possibilities through numerical simulations.  What does
the observed MDF and the luminosity width of the SGB associated with the
faintest TO tell us about the SFH?  The third, and final, subsection will
examine the thin RGB of Carina to ascertain the extent to which it is the 
outcome of a fortuitous alignment of the ages and metallicities of the stars
that populate this feature.

\subsection{Turnoff and HB Stars in Carina}
\label{subsec:TO}

In the upper panel of Figure 5, the CMD of M\,92 has been transposed to the 
[$(B-V)_0,\,M_V$]-diagram assuming (as adopted previously; see Fig.~3)
$E(B-V) = 0.019$ and $(m-M)_V = 14.75$.  This apparent distance modulus is
obtained if a ZAHB for $Y = 0.25$ and [Fe/H] $= -2.40$ (the solid curve)
is fitted to the lower bound of the distribution of the cluster HB stars (as
shown).  To obtain this result, the synthetic $B-V$ colors along the ZAHB were
adjusted to the red by 0.016 mag, which is obtained from our $\delta(B-V)$
tables if [Fe/H] $= -2.40$ and $\teff \ge 6750$~K.  Interestingly, V14b (see
their Fig.~5) found that the same ZAHB provided an equally good fit to the HB
of M\,92 on the [$(V-I_C)_0,\,M_V$]-diagram, but without having to apply any
adjustment of the synthetic colors (from CV14).  Essentially the same thing was
found by V13 (see their Fig.~11) in a study of
the nearly equivalent CMD of M\,92 that they constructed from
the {\it HST} $F606W,\,F814W$ observations that were obtained by Sarajedini
et al.~(2007) --- though the stellar models employed in the 2013 investigation
assumed a somewhat different metals mixture.  [While the main cause of
such minor inconsistencies is difficult to determine, it should not be a
surprise that they exist given that, e.g., on-going improvements to model
atmospheres and the line lists used in synthetic spectra are bound to affect
the magnitudes derived from some filter bandpasses more than others.  In
particular, $B$ is affected by line blanketing more than $V$ and $I_C$;
consequently, $B-V$ will be a more challenging color index to model than
$V-I_C$.]

\begin{figure}[t]
\plotone{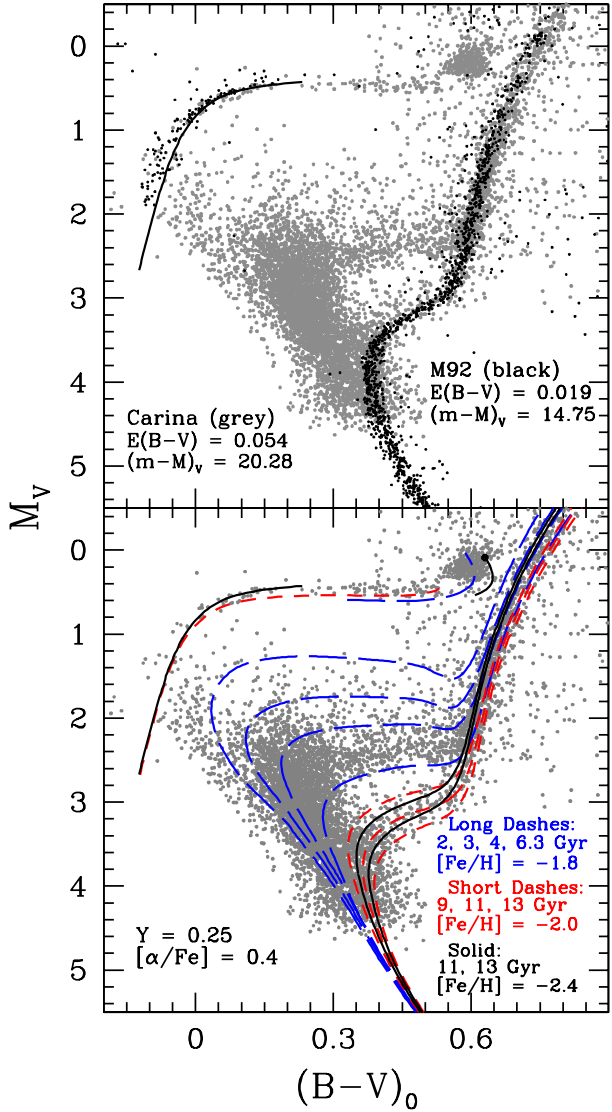}
\caption{{\it Upper panel:} Superposition of the M\,92 CMD onto that of
 Carina assuming the indicated values of the reddening and apparent distance
 moduli.   The solid curve represents a ZAHB for [Fe/H] $= -2.40$,
 [$\alpha$/Fe] $= +0.40$, and $Y=0.25$.  {\it Lower panel:} Overlay of
 isochrones for the indicated chemical abundances and ages, along with fully
 consistent ZAHB loci, onto the CMD of Carina.  Only the red end of the ZAHB is
 shown for [Fe/H] $= -1.8$ and [$\alpha$/Fe] $= +0.4$ (long dashes), as well as
 the one for [Fe/H] $= -1.4$ and [$\alpha$/Fe] $= 0.0$ (the curve with a small
 filled circle attached to the bright end).}
\end{figure}

If the reddening in the direction of Carina is $E(B-V) = 0.054$ (Schlafly \&
Finkbeiner 2011), the adoption of $(m-M)_V = 20.28$ results in the superposition
of the Carina and M\,92 CMDs shown in the top panel.  In this case, the cluster
SGB is coincident with the location of the densest concentration of Carina
subgiants that are associated with the faintest TO, implying a similar age if
the latter have the same metallicity.  However, judging from the very different
luminosity widths of their respective SGBs, the oldest stellar population in
Carina must span a much wider range in age or metallicity, or both, than the
stars residing in M\,92.  (These indications from photometry have, of course,
been confirmed by several spectroscopic studies; see \S~\ref{sec:intro}.)  The
assumed reddenings and distance moduli also imply that the RGB and HB of M\,92
lie along the blue edges of the galactic distribution of giants and horizontal
branch stars, respectively, which is consistent with M\,92 being more metal-poor
than the majority of the stars in Carina.  Encouragingly, the adopted apparent
modulus implies a true distance modulus of $(m-M)_0 \approx 20.11$, which is in
very good agreement with recent determinations based on near-IR magnitudes of
the RGB tip (20.08--20.12, Pietrzy\'nski et al.~2009), on the RR Lyrae variables,
(20.09, Coppola et al.~2013), and on the dwarf Cepheids in Carina (20.17, Vivas
\& Mateo 2013).

The location of the ZAHB for [Fe/H] $= -2.4$ relative to the blue HB stars in
Carina is more easily seen in the bottom panel, along with the same isochrone
(for 13 Gyr) that matches the TO and SGB of M\,92.  This isochrone lies below
most, but not all, of the faintest subgiants in the dSph: a better match to
the least luminous SGB stars is clearly provided by a 13 Gyr isochrone for
[Fe/H] $= -2.0$.  However, this is no more than suggestive, given that the
assumed value of [$\alpha$/Fe] in these models is too high (recall Fig.~1a), and
because a similar fit could be obtained using an even more metal-rich isochrone
for a somewhat younger age.  Still, these computations give the impression that
Carina contains stars at least as old as those found in the most metal-poor GCs
($\sim 13$ Gyr).  In addition, they illustrate that the luminosity width of
Carina's SGB is consistent with a range in age $\gta 4$ Gyr, at a fixed [Fe/H]
(note the separation between the 9 and 13 Gyr isochrones for [Fe/H] $= -2.0$),
or a variation in metallicity, at a given age, that is much larger than 0.4 dex
(note that the 13 Gyr isochrones for [Fe/H] $= -2.0$ and $-2.4$ enclose only a
small fraction of the observed subgiants).

\begin{figure}[t]
\plotone{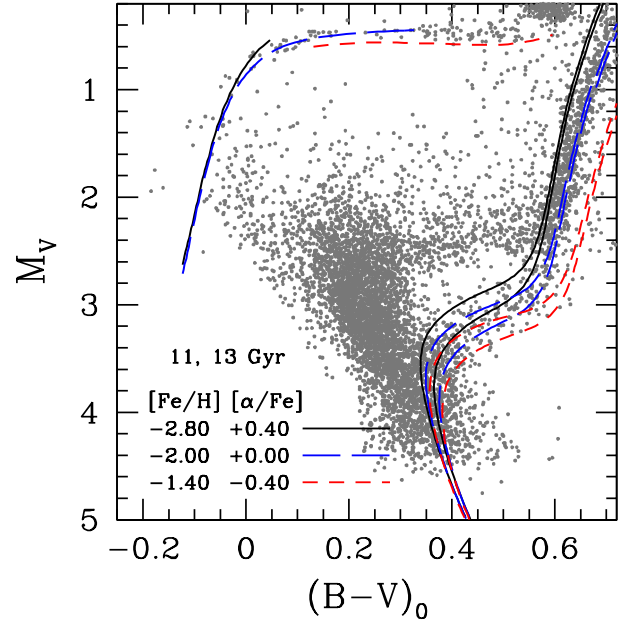}
\caption{Similar to the bottom panel in the previous figure, except that
 isochrones and ZAHBs for values of [Fe/H] and [$\alpha$/Fe] that are more
 consistent with spectroscopic determinations (see Fig.~1a) are compared with
 the Carina CMD.}
\end{figure}

The other isochrones which have been plotted indicate that little or no star
formation occurred in a 2--3 Gyr period prior to the major SF burst that
happened about 6 Gyr ago, and that there {\it may} have been a lull in the SF
activity $\sim 2.5$ Gyr ago as there is some (albeit marginal) indication of
enhanced numbers of MS stars at both younger and somewhat older ages.  It is
worth mentioning that, if isochrones for $Y = 0.25$, [Fe/H] $= -1.4$, and
[$\alpha$/Fe] $= +0.13$ (parameters that are also consistent with the
[$\alpha$/Fe] vs.~[Fe/H] relation shown in Fig.~1a) were compared with the
Carina CMD, we would obtain an age of 5.9 Gyr for the oldest of the
intermediate-age stars (as compared with 6.3 Gyr if they have [Fe/H] $= -1.8$
and [$\alpha$/Fe] $= +0.4$; see Fig.~5).  The metallicity dependence of this
age estimate is thus quite small --- because the change in the absolute oxygen
abundance between these two cases is only 0.13 dex.  As discussed by V12,
isochrones for the same oxygen abundance (i.e., [O/H]),
but different [Fe/H], will predict
nearly identical TO luminosity vs.~age relations (especially at low
metallicities, where the contributions of the metals to the opacities in the
interiors of stars are small). 

Aside from a 1--2 Gyr reduction in the absolute ages, these results are
quite similar to those reported by, e.g., Hurley-Keller et al.~(1998).  The
use of modern isochrones that take diffusive processes, recent advances in
nuclear reaction rates, etc., into account can be expected to have some impact
on age determinations, even if the same distances and chemical abundances are
assumed.  It is perhaps especially encouraging that current stellar models are
able to reproduce the morphologies and the locations of the different
photometric sequences in Carina very well once relatively small adjustments are
made to the predicted $B-V$ colors.  Something worth mentioning is that, as
shown many years ago by, e.g., Castellani \& Degl'Innocenti (1995, and
references therein), isochrones for young ages and very low metallicities have
relatively smooth turnoffs, in stark contrast with the pronounced blueward hook
features that are characteristic of their counterparts at near solar abundances.
In a qualitative sense, our 2--4 Gyr isochrones look very similar to those
computed $\gta 20$ years ago.  (Because the model grids used in this study are
limited to ages $\gta 5$ Gyr, the remainder of this paper will concentrate on
just the oldest stellar populations of Carina.)

Figure 6 illustrates the superposition on the Carina CMD of 11 and 13 Gyr
isochrones for metallicities that span nearly the full ranges in [Fe/H] and
[$\alpha$/Fe] that have been derived spectroscopically.  (Fully consistent
ZAHBs, to be discussed near the end of this subsection, are also shown.)  In
confirmation of the remarks made above concerning the oxygen abundance, the
isochrones for [Fe/H] $= -2.0$ and [$\alpha$/Fe] $= 0.0$ predict very similar
turnoff and SGB luminosities as those for the same ages but for [Fe/H] $= -2.40$
and [$\alpha$/Fe] $= +0.4$ (see the previous figure).  Both assume exactly the
same value of [O/H] and, as a close inspection of both plots reveals, their
locations relative to the observed TO and SGB stars in their vicinity are nearly
the same.  However, the RGB portions of these isochrones are obviously not
coincident, but that can be attributed to the differences in the abundances of
the heavier metals, notably the iron-peak elements.  Varying the oxygen
abundance does not affect the location of the RGB at low metallicities (see
V12). 

Similarly, considering the isochrones for [Fe/H] $= -2.8$ and [$\alpha$/Fe]
$= +0.4$, on the one hand, and those for [Fe/H] $= -2.0$ and [$\alpha$/Fe]
$= 0.0$, on the other, it is readily understood that it is the 0.4 dex
difference in [O/H] which is primarily responsible for the $\sim 1$ Gyr
difference in age at a given TO luminosity (see Fig.~6).  Although the
abundances of the iron-peak elements differ by 0.8 dex, and those of the other
$\alpha$-elements (Ne, Mg, Si, $\ldots$) differ by 0.4 dex, these variations
mainly cause a small shift in the model $\teff$\ scale due to their effects on
the atmospheric opacities (see V12).  The implication of these results is that
the age at a given TO luminosity varies quite weakly with position along the
sloped parts of the [$\alpha$/Fe] vs.~[Fe/H] relation shown in Fig.~1a.
Conversely, a given subgiant thickness would imply a narrower range in [Fe/H],
at a common age, if [$\alpha$/Fe] $=$ {\it constant} than if [$\alpha$/Fe] and
[Fe/H] are anti-correlated (as observed). 

This undoubtedly helps to explain why Bono et al.~(2010) argued that the oldest
TO/SGB stars in Carina span a significantly smaller range in [Fe/H] than that 
derived from spectroscopic work.  They reached this conclusion as the result
of (i) comparing the fiducial sequences for a few of the Galactic GCs, which
have [$\alpha$/Fe] $\approx 0.4$, with the Carina CMD, and (ii) examining
synthetic CMDs that were derived from isochrones for [$\alpha$/Fe] $= 0.0$ and
a wide range in [Fe/H].  Both of these avenues assumed, either directly or
indirectly, constant [$\alpha$/Fe]; in which case, isochrones of similar age
but different [Fe/H] (and [O/Fe], since $\delta$[O/H] $=$ $\delta$[Fe/H]) will
predict a larger difference in their TO/SGB luminosities than one expects if
[$\alpha$/Fe] decreases with increasing [Fe/H].  What is of critical 
importance in this discussion is how [O/H] (which is equivalent to the absolute
O abundance, if the models being compared assume the same solar abundances)
varies with [Fe/H].  

Perhaps the most compelling argument presented by Bono et al.~(2010) is that,
as demonstrated in their CMD simulations, the component of Carina's HB that is
populated by its oldest stars is too thin to be consistent with the range in
[Fe/H] that has been derived in spectroscopic studies.  Indeed, this is also
suggested by the ZAHB models that appear in the bottom panel of Fig.~5.  Most
of the old core He-burning stars (i.e., those populating the nearly flat portion
of the HB at $(B-V)_0 \lta 0.55$), are brighter than the ZAHB for [Fe/H]
$= -2.0$ (the short-dashed curve) and fainter than the one for [Fe/H] 
$= -2.4$ (the solid curve).  Moreover, there is no evidence at all for
old HB stars with [Fe/H] $\gta -1.8$ as they should lie below the ZAHB
represented by the long-dashed curve.  

Before moving to the next figure, note that the last of these three ZAHBs, which
has been extended to a mass that is predicted for a 2 Gyr RGB tip star, reaches
close to the maximum luminosity of the HB clump (as shown).  Encouragingly, a
star of the same [Fe/H] that has an RGB tip age of 6 Gyr (approximately the age
of the oldest IA stars) is predicted to reach its ZAHB location at close to the
minimum luminosity of the clump (at the location of the long-dashed
curve).  According to our models, and consistent with the observed MDF to
within its uncertainties, few of the observed HB clump stars are predicted to
have [Fe/H] $> -1.4$ (which has been assumed, together with [$\alpha$/Fe]
$= 0.0$, in the computation of the ZAHB to which a small filled circle has been
attached at its bright end), since this ZAHB matches the red edge of the
distribution of HB clump stars.  (These abundances are consistent with the
variation of [$\alpha$/Fe] with [Fe/H] given by the solid curve in
Fig.~1a.)  To reinforce this point, a ZAHB for [Fe/H] $= -1.2$
and [$\alpha$/Fe] $=0.0$ would have been located well to the red of the HB
clump, almost intersecting the observed RGB, had we chosen to plot this case.
(There may well be some stars in our CMD that have [Fe/H] $\sim -1.2$, in
accordance with spectroscopic findings, but the small minority of such stars
presumably does little more than contribute to the scatter in the vicinity of
the HB.)

The benefits of being able to compare observations with stellar models {\it for
the observed chemical abundances} is further revealed by the ZAHBs that have
been plotted in Fig.~6.  They show that the small luminosity width of Carina's
old HB does not preclude the presence of stars with [Fe/H] values that vary by
more than 1 dex, {\it provided} that [$\alpha$/Fe] varies with [Fe/H] in
approximately the observed way (i.e., following the trend given by the
dashed line in Fig.~1a).  The ZAHB for [Fe/H] $= -1.4$ and [$\alpha$/Fe]
$= -0.4$ provides quite a good fit to the faintest of the cluster
HB stars that are fainter than the clump stars and that have $0.35 \lta (B-V)_0
\lta 0.55$.  This is not unexpected since, as reported by VandenBerg et
al.~(2000), the luminosity of a ZAHB is quite a strong, inverse function of
[$\alpha$/Fe], at fixed values of [Fe/H] and $Y$.\footnote{Work is
in progress by D.A.V.~to produce ZAHB loci for all of the chemical abundance
choices considered by V14a, and we leave a detailed discussion of those ZAHBs
to the forthcoming paper.  However, it may be of some interest to note
that, despite the revisions which have been made to the Victoria evolutionary
code over the past 15 years, current ZAHB models predict very similar
luminosities for RR Lyrae stars as those presented by VandenBerg et al.~(2000).
Whereas the latter neglected diffusion physics and assumed a primordial helium
abundance corresponding to $Y = 0.235$, which was the favored value at that 
time, current Victoria models take the gravitational settling of helium into
account while assuming the current best estimate of the primordial value of $Y$
($\approx 0.250$; e.g., Komatsu et al.~2011).  Fortuitously, this increase in
$Y$ largely compensates for the effects of diffusion on the properties of RGB
tip stars (primarily the envelope helium abundance) that control the predicted
luminosity of the HB.}

To be sure, the distribution of the stars relative to the three ZAHBs that have
been plotted in Fig.~6 seems anomalous.  Most of the observed HB stars are
fainter and redder than the ZAHB for [Fe/H] $= -2.0$ and [$\alpha$/Fe] $= 0.0$,
implying that they are more metal rich, which is obviously in conflict with the
MDF.  A possible solution to this dilemma is that the oxygen abundance is
significantly higher than [O/Fe] $= +0.4$ in (at least) the most metal-deficient
stars in Carina, and that it remains offset from the [$m$/Fe] values of the
other $\alpha$-elements by $\sim 0.2$ dex (or more?) as [$\alpha$/Fe] decreases
with increasing [Fe/H].  For instance, [O/Fe] $\gta +0.6$ is generally found in
Milky Way halo stars (e.g., Fabbian et al.~2009, Ram\'irez et al.~2012, V14b),
whereas a smaller enhancement ([$m$/Fe] $= 0.3$--0.4) is generally found for
Mg, Si, Ca, Ti, etc.~(see, e.g., Cayrel et al.~2004).  At low metallicities, a
ZAHB is stretched to redder colors and fainter luminosities over a considerable
fraction of its length as the absolute oxygen abundance is increased (see,
e.g., VandenBerg \& Bell 2001; their Fig.~6).  (At higher [Fe/H], enhanced
[O/H] mainly causes a redward displacement of just the red end of a ZAHB.)
Thus, higher oxygen could shift the ZAHB loci for [Fe/H] $\lta -2.0$ in just
the way that is needed to reconcile them with the locations of a signficant
fraction of the reddest stars in the ``old" HB component of Carina.
Unfortunately, little seems to be known about the [O/H] values in Carina
stars, and it must be left for future work, both observational and theoretical,
to investigate these speculations.

Figure 6 also suggests that the faintest SGB stars are old, iron-rich,
$\alpha$-poor stars.  (The delay time between SF and the onset of SN Type Ia
explosions is not necessarily an argument against the near coevality of stars
with very different [$\alpha$/Fe] values, as it can be as short as $\sim
100$--500 Myr or as long as a few Gyr, depending on, among other things, whether
the star formation rate is high or low, respectively; see the review by Maoz et
al.~2014.)  It is possible that some of them are field stars, though we have
tried to remove such objects.  Alternatively, it may be that the actual [O/Fe]
values are higher than those implied by the assumed values of [$\alpha$/Fe].
This would have the effect of reducing the predicted age at a given TO
luminosity, or equivalently, of shifting isochrones for a fixed age and [Fe/H]
to higher $M_V$.  That is, the faintest SGB stars could be very old stars with
low iron, and relatively high oxygen, abundances (that evolve to red
ZAHB locations).  To provide a limited investigation of the consequences of
this possibility for the oldest TO/SGB stars in Carina, some of the simulations
presented in the next subsection will make use of the model grids for
[$\alpha$/Fe] $= +0.4$ and high values of [O/Fe] that were employed in the
recent study of several UFD galaxies by Brown et al.~(2014).

\subsection{Simulations of the SFH of the Oldest Stellar Populations}
\label{subsec:sim}

\begin{figure}[t]
\plotone{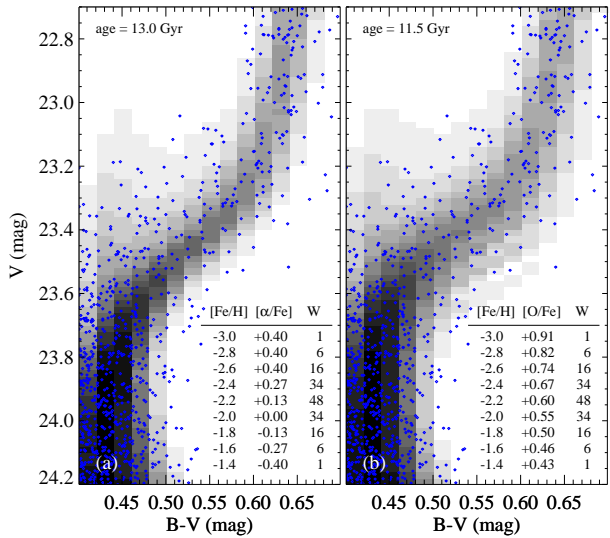}
\caption{Best-fit model (grey shading) for the oldest Carina stars (blue 
points), assuming a single burst of star formation and the [Fe/H] distribution 
of Figure~1b. The relative weights ($W$) are also specified in the inset table 
for each panel.  The $\alpha$-varying model produces a much narrower SGB
than observed ({\it left panel}).  The O-varying model (with [$\alpha$/Fe]
otherwise held at +0.4) produces a somewhat broader SGB that is still 
narrower than observed ({\it right panel}).}
\end{figure}

\epsscale{0.8}
\begin{figure*}[h]
\plotone{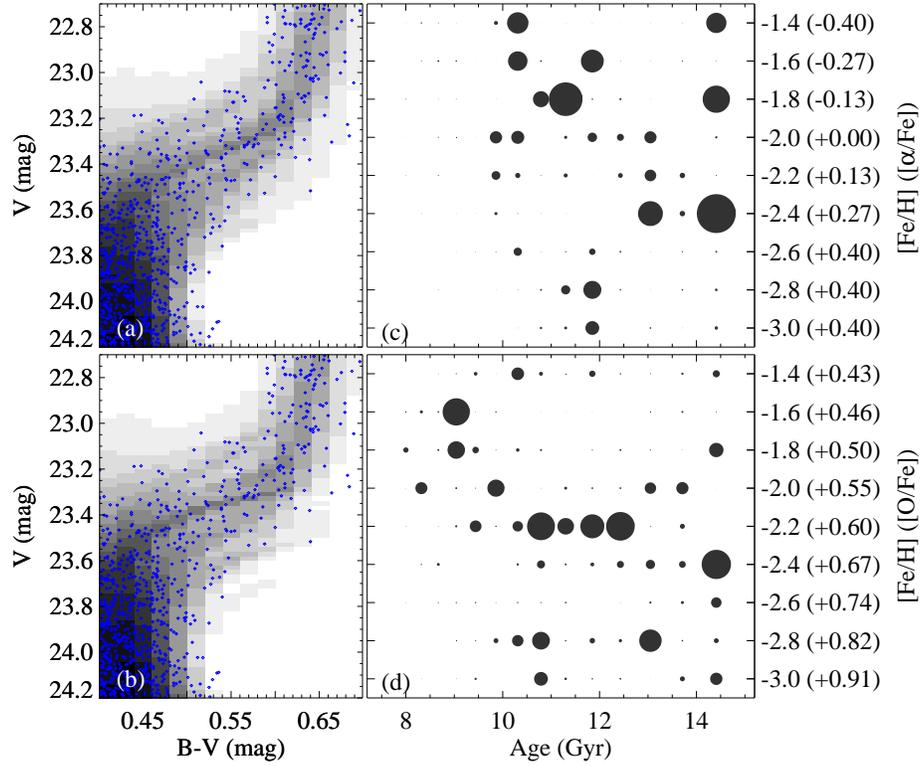}
\caption{Best-fit model (grey shading) for the oldest Carina
population (blue points), assuming an extended SFH.  
Metallicities are constrained to the [Fe/H] range in Figure~1b, with both
an $\alpha$-varying model ({\it panel a}) and an O-varying model
({\it panel b}).  Associated with each synthetic CMD, we show the
best-fit SFH ({\it panels c and d}), with the area of each circle
indicating the weight at that point in the grid of age and [Fe/H].}
\end{figure*}
\epsscale{1.0}

To explore the SFHs that are consistent with the oldest populations
in Carina under different metallicity assumptions, we have constructed
several synthetic CMDs in the $B$ and $V$ bands, employing the same
calibrated isochrone library described above.  Two different SFH
scenarios have been explored: a short burst of star formation with an
idealized MDF (Fig.~1b), and an extended period of star formation lasting
several Gyr, with only loose constraints on the MDF.  In each
of these two SFH scenarios, the relationships for [$\alpha$/Fe] vs.~[Fe/H]
and [O/Fe] vs.~[Fe/H] were also varied.  A linear combination of synthetic
CMDs, each representing a particular age and metallicity, was used to
fit the observed CMD.  These fits were performed via the minimization
of the Poisson maximum likelihood statistic of Dolphin (2002), and the
quality of the best fit was evaluated by comparing the best-fit score to
the distribution of such scores arising from Monte Carlo realizations of
the photometric data.

Usually, synthetic CMDs are constructed using the results of
artificial star tests.  Such tests characterize the photometric
uncertainties and incompleteness as a function of position within the
CMD, allowing each point along an isochrone to be scattered in the
synthetic color-magnitude plane.  Due to the heterogeneous nature of the
observations studied here, artificial star tests are impractical, but
the photometry of each star in the catalog comes from many individual
exposures, such that the photometric uncertainties are well understood.
Furthermore, for the CMD region being simulated, the photometric errors
are small ($<$0.05~mag); consequently, the completeness should be near
100\%.  For these reasons, we used the photometric errors from the
catalog to define a Gaussian scattering kernel as a function of position
within the CMD.  

The synthetic CMDs were populated using a Salpeter (1955) initial mass
function (IMF), although the small mass range of stars in the fitting
region ($< 0.1 {{\cal M}_\odot}$) makes the assumed IMF unimportant.
We assumed a binary fraction of 14\% (Minor 2013), but the results
would not change significantly if we were to adopt a value closer to
50\%, which is typically found in other dwarf galaxies.  To fit the
SFH of the oldest population, we isolated the faintest SGB stars in 
the catalog, along with its neighboring stars in the vicinity of the
MSTO and the lower RGB (see Figure~7).

We begin with a simple model in which (i) the star formation is
assumed to occur in a single burst, and (ii) the variation in the
number of stars as a function of [Fe/H] corresponds to the MDF shown in
Fig.~1b.  However, we have explored two distinct assumptions regarding
[$\alpha$/Fe] and [O/Fe].  The first is the ``$\alpha$-varying'' model,
where [$\alpha$/Fe] increases with decreasing [Fe/H] (as in Figure~1a).
The second is the ``O-varying'' model, where [O/Fe] increases with
decreasing [Fe/H], while all other $\alpha$ elements are held at a
constant [$\alpha$/Fe]~=~+0.4 (as typically found in old, very metal-poor
populations).  The adopted relation between [O/Fe] and [Fe/H] is the
same as that assumed by Brown et al.~(2014), which is based on data
presented by Frebel (2010).

\epsscale{0.8}
\begin{figure*}[t]
\plotone{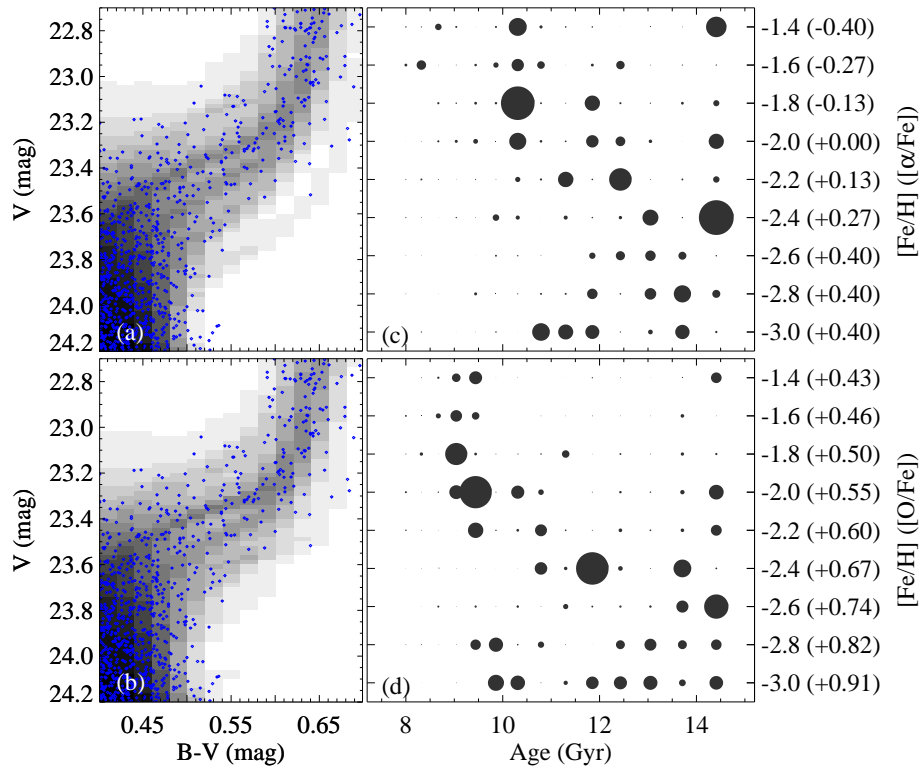}
\caption{As in Figure~8, but for an assumed distance modulus that
is 0.1~mag larger.}
\end{figure*}
\epsscale{1.0}

Figure~7a shows the best-fit burst with the $\alpha$-varying model.
The simulated SGB is significantly narrower than the observed SGB,
resulting in a rather poor fit ($\chi_{\rm eff}^2$=1.8).  Figure~7b shows
the best-fit burst in the case of the O-varying model.  The simulated
SGB is broader, but still not as broad as that observed, again producing an
unsatisfactory fit ($\chi_{\rm eff}^2$=1.4).  Although not shown here,
we note that a model in which all of the $\alpha$-elements, including
oxygen, have a constant enhancement (specifically, [$\alpha$/Fe] $= +0.4$)
for all [Fe/H] values looks quite similar to that illustrated in
Figure~7b.  Both of these fits are ruled out at more than $6\,\sigma$.
Thus, if the MDF in the oldest population of Carina is similar to that
shown in Figure~1b, it is unlikely to be a single-age population.

We now turn to a model with a more complex SFH, shown in Figure~8.
Instead of a single burst, the age distribution is only restricted to 
the range of 8 to 14.4~Gyr.  The MDF is constrained to iron abundances
$-3.0 \le$ [Fe/H] $\le -1.4$, as in the case of the single burst models,
but it is permitted to take any form within this range.  (These fits were
performed using the StarFISH code of Harris \& Zaritsky 2001.)  Our
results for an $\alpha$-varying model and an O-varying model are shown
in the upper and lower panels, respectively.  For both cases,
$\chi_{\rm eff}^2 = 1.2$ (rounded to one decimal place), but the
O-varying fit is the slightly better one.  However, although nearly the
same value of the goodness-of-fit statistic is obtained, the derived
weighting of the isochrones in Fig.~8d resembles the observed MDF more so
than that shown in Fig.~8c.  In particular, the former simulation predicts
more stars with [Fe/H] $\sim -2.2$ (recall Fig.~1b) than in lower or higher
metallicity bins, whereas the opposite is found in the latter simulation.

It may be recalled that the results presented in Fig.~8c were anticipated
by the overlays of isochrones to the Carina CMD discussed previously.  As
shown in Fig.~6, isochrones for high ages {\it and} [Fe/H] values are
apparently needed to explain the faintest SGB stars if the assumed
$\alpha$-element abundances follow the adopted relation between
[$\alpha$/Fe] and [Fe/H].  As indicated by the simulation results in
Fig.~8d, this is not the case if the stellar models allow for high oxygen
abundances.  Also worth mentioning is the fact that significantly higher
ages, at a fixed metallicity, would have been found had we neglected
diffusive processes (see, e.g., VandenBerg et al.~2002, their Fig.~2).
(However, if such physics is ignored, it would become especially difficult
to reconcile stellar ages with the age of the universe; see V14b.)

Although the fits are not perfect, both of the cases considered in Fig.~8
demonstrate that the oldest population in Carina can be reproduced if the
star formation epoch lasted a few Gyr.  
%Whether stars span a similar range
%in age at most [Fe/H] values (Fig.~8c), and/or their median ages tend to
%decrease somewhat with increasing metallicity (Fig.~8d), depends quite
%sensitively on how [O/Fe] varies with [Fe/H].  Obviously, the distribution
%and variation of stellar ages also depends on the assumed distance modulus.
Although there is a general trend of decreasing age at increasing
metallicity, the particular distribution of the points in the age-metallicity
plane depends quite sensitively on how [O/Fe] varies with [Fe/H], and the
distance modulus assumed.
A shorter distance would be more problematic than a longer distance
because it would imply that all of the stars (notably those on the SGB)
have fainter $M_V$ values, thereby making them older.  As a consequence,
the weighting of the isochrones found for the O-varying model (Fig.~8d) 
would move in the direction of that obtained for the $\alpha$-varying 
model (Fig.~8c), and the results for the latter case would be skewed to
even older ages and higher [Fe/H] values (in greater conflict with the
observed MDF).

Figure~9 shows that the opposite trend is found if a slightly larger distance
modulus (by 0.1 mag) is assumed.  Comparing, e.g., Fig.~9c with Fig.~8c
reveals that the distribution and weighting of the points has been shifted
to lower [Fe/H] values.   That is, the models predict larger {\it fractions}
of older, more metal-poor stars if the distance to Carina is 0.1 mag larger
than our adopted (ZAHB-based) modulus.  (The same thing, which may seem
counter-intuitive, is implied by the differences between Figs.~9d and 8d.)  As
a consequence, the $\alpha$-varying model provides a more agreeable fit to the
observations, while the O-varying model seems to be somewhat less satisfactory.
However, the main value of the simulations presented in Figs.~8 and 9 is to
illustrate how they are affected by parameter variations.  To obtain accurate
and robust age determinations, it is clearly of paramount importance to have
reliable abundances of the individual $\alpha$-elements (especially O, but 
also Mg and Si; see V12) so that their variations with [Fe/H], as well as the
observed star-to-star scatter of [$m$/Fe] at a fixed iron abundance, can be
taken into account when simulating the observed CMD of Carina.

\subsection{The Thin RGB of Carina}
\label{subsec:RGB}

The fact that Carina has a very narrow RGB is unexpected given the wide ranges
in ages and metal abundances encompassed by its stellar populations.  As Venn
et al.~(2012) have remarked in their recent study of this dSph, it is possible
that the giants ``have a fortuitous alignment in the age-metallicity degeneracy,
such that the older metal-poor stars overlie the metal-enhanced
intermediate-age stars".  Although this overlap is undoubtedly happening, to
what extent {\it does} the observed RGB constrain the properties of Carina's
stellar populations?  In what follows, we will endeavor to answer this question
using several of the same isochrones that seem to provide quite good fits to
the TO and HB observations.

\begin{figure}[t]
\plotone{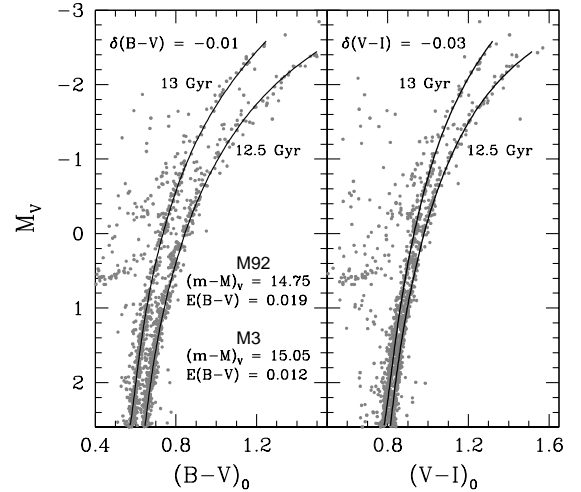}
\caption{Fits of the RGB segments of 13.0 Gyr, [Fe/H] $= -2.40$ and 12.5
 Gyr, [Fe/H] $= -1.55$ isochrones to the giant branches of M\,92 and M\,3,
 respectively.  (M\,92 has a bluer RGB than M\,3.)  Both isochrones assume
 $Y = 0.25$ and [$\alpha$/Fe] $= 0.40$ and both were adjusted by $\delta(B-V)
 = -0.01$ and $\delta(V-I) = -0.03$ mag in order to provide the best overall
 fits to the cluster observations.}
\end{figure}

\begin{figure*}[t]
\plotone{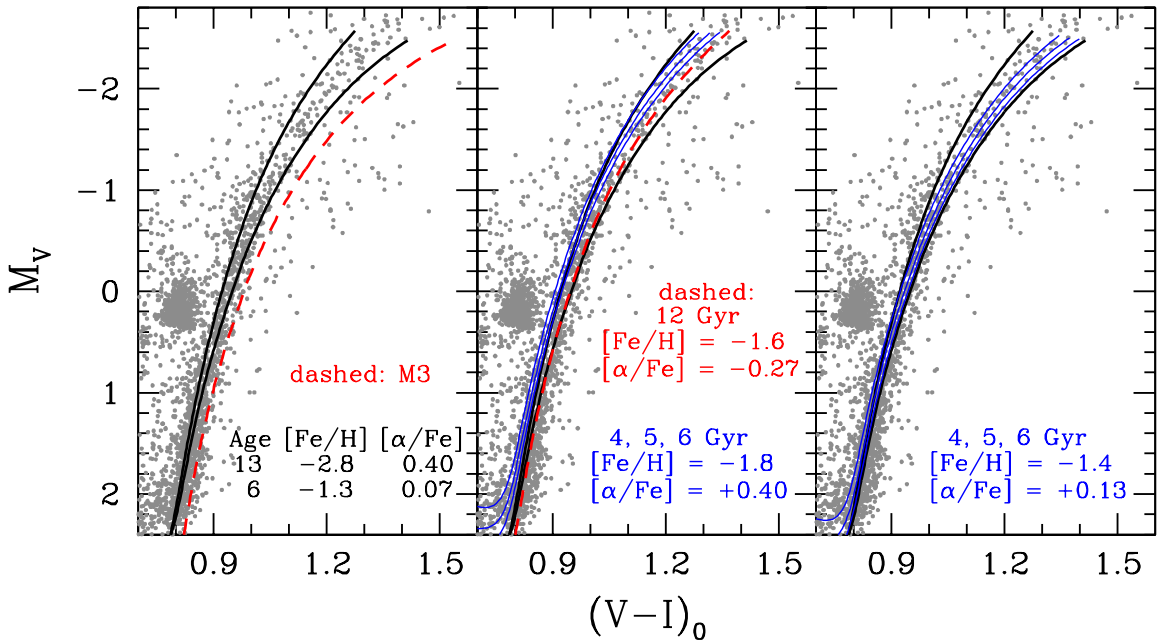}
\caption{{\it Left-hand panel:} Overlay of the RGB portions of three
 isochrones onto the Carina CMD.  The bluer of the two solid curves assumes
 [Fe/H] $= -2.40$, [$\alpha$/Fe] $= 0.40$, and an age of 13 Gyr; the other one
 was generated for [Fe/H] $= -1.30$, [$\alpha$/Fe] $= 0.07$, and an age of 
 6 Gyr.  The dashed curve is the best-fit isochrone to the M\,3 RGB (see the
 previous figure).  {\it Middle panel:} As in the left-hand panel, except
 that the dashed curve has the indicated properties and the thin solid curves
 represent the RGB portions of 4, 5, and 6 Gyr isochrones for [Fe/H] $= -1.8$
 and [$\alpha$/Fe] $= 0.40$.  {\it Right-hand panel:} The same 6 and 13
 Gyr isochrones that appear in the left-hand and middle panels have been
 replotted, along with three isochrones for the indicated ages and chemical
 properties (the thin solid curves).  The adopted reddening and distance
 modulus of Carina are the same as in previous figures, and we have assumed
 that $E(V-I) = 1.244\,E(B-V)$ (Casgrande \& VandenBerg 2014).  The predicted
 colors along the isochrones were adjusted by $\delta(V-I) = -0.03$ mag prior
 to plotting them.} 
\end{figure*}

Fortunately, the predicted morphologies of current Victoria-Regina isochrones
reproduce the shapes of GC giant branches very well when standard reddenings
(from Schlafly \& Finkbeiner 2011), distances based on ZAHB models that
satisfy RR Lyrae constraints (see V13, V14b), and metallicities close to those
derived by, e.g., Kraft \& Ivans (2003) and Carretta et al.~(2009a) are
assumed.  As shown in the left-hand panel of Figure 10, the isochrones for the
same ages and chemical abundances that were fitted to the turnoff observations
of M\,3 and M\,92 in \S~\ref{subsec:calib}, on the assumption of the cluster
properties that are given explicitly in Fig.~3, provide a close match to their
RGB populations on the [$(B-V)_0,\,M_V$]-diagram.  To accomplish this, the
$B-V$ colors from the CV14 transformations had to be corrected by just $-0.01$
mag, independently of gravity and $\teff$.  The right-hand panel shows that
equally good agreement can be obtained on the [$(V-I)_0,\,M_V$]-diagram; but in
this case, the required color adjustment turns out to be $\delta(V-I) = -0.03$
mag.  (The origin of the small zero-point offsets is not known, but they can be
reasonably attributed to small photometric errors or minor problems with, e.g.,
the CV14 color transformations, the model $\teff$\ scale, and/or the assumed
cluster properties.)

What is especially encouraging is that exactly the same $\delta(B-V)$ and
$\delta(V-I)$ color offsets are needed for both M\,3 and M\,92, despite the
difference of nearly 1 dex in their [Fe/H] values.  This implies that models
which are fitted to the giant branch of Carina should be trustworthy in an
absolute sense if we simply apply the same color adjustments to them.  In
effect, we have calibrated the RGBs of the Victoria-Regina isochrones using
M\,92 and M\,3.  (As shown by V14a, these computations are equally successful
over a much broader range in [Fe/H].  The main difficulty, as also reported by 
V13, is that somewhat different offsets are needed to fit the MSTO observations
than to match the location of cluster RGBs; i.e., the predicted difference in
color between the TO and the lower giant branch differs from the observed
difference by a small amount.  However, this is not of particular concern
because, as noted at the end of the previous paragraph, such discrepancies
could easily be due to relatively minor deficiencies in some of the more
uncertain aspects of the stellar models, the assumed chemical compositions,
etc.)

It is already apparent in Fig.~10 that a difference of 0.85 dex in [Fe/H]
implies a much thinner RGB on the [$(V-I)_0,\,M_V$]-diagram, especially at $M_V
\gta 0.0$, than on the [$(B-V)_0,\,M_V$]-plane.  This is the expected
consequence of the fact that the $B-V$ color index has a much stronger
metallicity dependence than $V-I$.  Because the latter will also be less
affected by variations in the metals mixture, it should be more straightforward
to interpret the fits of isochrones to the $VI$ photometry of Carina than to
CMDs based on $BV$ data.  Therefore, we will first examine the implications for
the observed RGB on the [$(V-I)_0,\,M_V$]-diagram of the isochrones that have
been compared with the TO observations in the previous section.  ($V-I$ colors
were not considered previously because the brightness of the night sky in $I$
makes it impossible to obtain ground-based photometry in this filter for the
faintest TO stars in Carina.)

Figure 11 compares the giant branches of several isochrones with our $VI$ data,
assuming that Carina has $E(B-V) = 0.054$ and $(m-M)_V = 20.28$, as adopted
throughout this investigation.  To deredden the $V-I$ colors, we have assumed
that $E(V-I) = 1.244\,E(B-V)$ (see CV14).  Since $> 95$\%
of the stars in the dSph appear to have [Fe/H] values between $-2.8$ and
$-1.3$ (according to the MDF in Fig.~1b), we have plotted (in all three panels, 
as thick solid curves) the RGBs of a 13 Gyr isochrone for [Fe/H] $= -2.8$,
[$\alpha$/Fe] $= +0.4$ and a 6 Gyr isochrone for [Fe/H] $= -1.3$, [$\alpha$/Fe]
$= +0.07$ (consistent with the relation between [$\alpha$/Fe] and [Fe/H] shown
in Fig.~1a).  It is apparent (see the left-hand panel) that these isochrones
bracket the upper-RGB populations quite well and that they predict an
exceedingly thin lower RGB.  The dashed curve represents the giant-branch
segment of the the same isochrone that was superimposed on the CMD of M\,3 in
Fig.~10.  This is significantly redder than the majority of the bright
giants in Carina, from which one would be inclined to conclude, in the absence
of any additional information, that the galaxy has few, if any, stars with
[Fe/H] values as high as $-1.55$.  However, we know that from spectroscopic
work that there are large numbers of giants with higher iron, but lower 
$\alpha$-element, abundances, and photometric studies have revealed the
existence of many young stars.  In accordance with standard stellar evolutionary
theory (e.g., see V14a), the most iron-rich giants in Carina are bluer than the
RGB of M\,3, at the same $M_V$, because the former are younger and $\alpha$-poor
(in an absolute sense) relative to their cluster counterparts.\footnote{It
should be kept in mind that it is the absolute abundances of the metals, not
[$m$/Fe] ratios, that are important for stellar models.}

\begin{figure*}[t]
\plotone{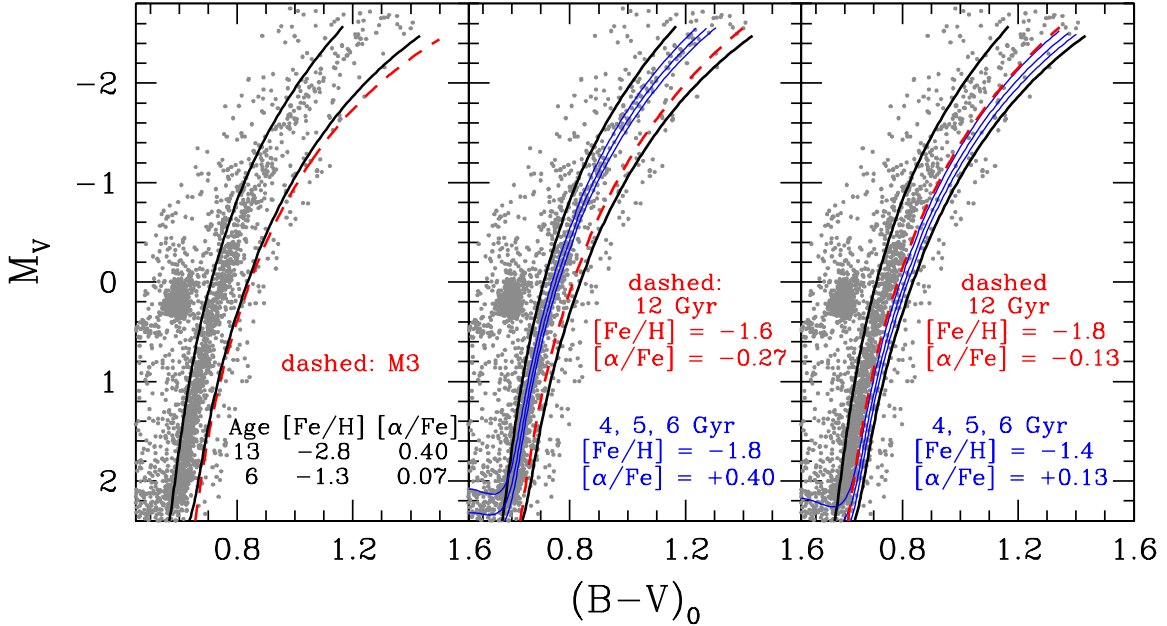}
\caption{As in the previous figure, except that the comparisons between theory
 and observations are made on the [$(B-V)_0,\,M_V$]-diagram and the isochrone
 colors were corrected by  $\delta(B-V) = -0.01$ mag.}
\end{figure*}

In the middle panel of Fig.~11, a 12 Gyr isochrone for [Fe/H] $= -1.6$ and
[$\alpha$/Fe] $= -0.27$ has been overlaid onto the Carina CMD, along with the
same two isochrones from the left-hand panel (the thick solid curves) and 4--6
Gyr isochrones for [Fe/H] $= -1.8$, [$\alpha$/Fe] $= +0.4$ (the thin solid
curves).  The separation in color at a fixed magnitude between the old $-2.8$
and $-1.6$ isochrones is quite small --- easily consistent with the observed
spread in RGB colors at a given luminosity (at least on this CMD).
Whereas the three isochrones for intermediate ages lie between those
for [Fe/H] $= -2.8$ and $-1.3$ at $M_V \lta -1$, they are located slightly to
the blue of the lowest metallicity isochrone at fainter absolute magnitudes.
Even so, the width of the RGB below the HB clump is still very narrow, and it
would remain so even if the range in [Fe/H] or age were expanded somewhat.  In
fact, the predicted RGB is actually a bit thinner than the observed one, which
may be a consequence of photometric errors, small as they are (see
\S~\ref{sec:obs}).  Note that old ($\sim 12$ Gyr) isochrones for [Fe/H] $< -1.6$
would overlie the 4--6 Gyr isochrones that have been plotted for [Fe/H]
$= -1.8$, which illustrates the well-known age-metallicity degeneracy.

Very similar comparisons between theory and observations are shown in the
right-hand panel; in this case, the intermediate-age isochrones are for [Fe/H]
$= -1.4$ and [$\alpha$/Fe] $= +0.13$.  Not unexpectedly, they lie just to the
blue of the 6 Gyr isochrone for [Fe/H] $= -1.3$ and [$\alpha$/Fe] $= +0.07$.
A similar set of 4--6 isochrones was computed for [Fe/H] $= -1.6$ and
[$\alpha$/Fe] $= +0.27$, which is also consistent with the [$\alpha$/Fe] 
vs.~[Fe/H] relation in Fig.~1a, and although we have chosen not to plot them
here, they lie approximately midway between those for [Fe/H] $= -1.8$ (middle
panel) and $-1.4$ (right-hand panel).  As a result, they would overlie the
densest concentration of Carina giants better than the latter, which is perhaps
to be expected since the peak in the MDF occurs at [Fe/H] $\sim -1.65$ (see
Fig.~1b).

In hindsight, it is not too much of a surprise that the Carina RGB is very
narrow on the [$(V-I)_0,\,M_V$]-diagram, given that the vast majority of its
stars appear to be more metal-rich than M\,92, more metal-poor than M\,3
(largely due to the low abundances of the $\alpha$-elements), and older than
$\sim 3$ Gyr (see Fig.~5).  Because nearly all of the giants in the dSph lie
to the blue of M\,3's RGB, (see the left-hand panel of Fig.~11), it is to be
expected that the thickness of the galactic RGB will be less than the separation
between the M\,92 and M\,3 giant branches, which amounts to only $\delta(V-I)
\approx 0.04$ mag at $M_V$ values that are fainter than the cluster HBs (see
Fig.~10).  Young ($< 3$ Gyr), [Fe/H] $\lta -1.8$ Gyr giants would be located on
the blue side of a 13 Gyr, [Fe/H] $= -2.8$ isochrone, but they do not appear to
be present, at least in significant numbers (see Fig.~11).  We conclude that
the thin RGB of Carina is completely consistent with the observed and derived
properties of its stellar populations, even if moderately large uncertainties
are taken into account.

However, $V-I$ colors are a better discriminator of differences in $\teff$
than of metallicity, so it is instructive to perform the same comparisons of
isochrones with the RGB of Carina as those just discussed, but on the
[$(B-V)_0,\,M_V$]-diagram.  Indeed, the analogous plot (see Figure 12) 
differs by more than one might have anticipated from the previous one, though
there are several similarities as well.  The most noticeable difference is that
the giants are not as centrally located between the two thick solid curves as
in Fig.~11 (compare the left-hand panels).  Indeed, the 6 Gyr isochrone for
[Fe/H] $= -1.3$ and [$\alpha$/Fe] $= +0.07$ is significantly redder than the red
edge of the distribution of giants, though a much better fit to the latter is
obtained by a 4 Gyr isochrone for the same metal abundances (judging from the
location of the isochrone for [Fe/H] $= -1.4$ and [$\alpha$/Fe] $= +0.13$ in
the right-hand panel).  In addition, the median RGB of Carina seems to be
matched quite well by 5--6 Gyr isochrones for [Fe/H] $= -1.8$ and [$\alpha$/Fe]
$= +0.4$ (see the middle panel).  Both of these isochrones were located just
to the left of the main concentration of bright giants on the
[$(V-I)_0,\,M_V$]-plane.

As found previously, a 13 Gyr isochrone for [Fe/H] $= -2.8$ and [$\alpha$/Fe] $=
+0.4$ provides a good fit to the blue edge of the distribution of Carina giants.
Why the same interpretation of the redder stars is not found from both $B-V$
and $V-I$ colors is not clear, but there are a few possible explanations for
this.  For one, it is possible that the [Fe/H] scale derived by Starkenburg et
al.~(2010), which has been assumed here (see Fig.~1b), is too metal-rich by
$\sim 0.15$--0.2 dex.  Such a shift, which is $\lta$ the differences in measured
GC metallicities over the years (e.g., see Carretta \& Gratton 1997, Kraft \&
Ivans 2003, Carretta et al.~2009a) would affect the horizontal separation
between the two isochrones in the left-hand panel of Fig.~12 considerably more
than in the case of the corresponding isochrones in Fig.~11.\footnote{If
anything, spectroscopic studies in the past few years have caused even more
consternation regarding {\it absolute} [Fe/H] determinations than earlier ones.
For instance, whereas Carretta et al.~(2009a) found that M\,15, M\,92, and
NGC\,5466 all have [Fe/H] values within the range of $-2.31$ to $-2.35$, Sobeck
et al.~(2011) and Roederer \& Sneden~(2011) have obtained [Fe/H] $\lta -2.6$
for M\,15 and M\,92, respectively, while Lamb et al.~(2015) have derived a
value of $-1.97$ for NGC\,5466.}  We see, for instance, that a 12 Gyr
isochrone for [Fe/H] $= -1.8$ provides a much better fit to the red edge of the
main distribution of Carina giants (right-hand panel) than the 12 Gyr isochrone
for [Fe/H] $= -1.6$ that has been plotted in the middle panel.  (If it is only
the iron abundance that changes, the derived value of [$\alpha$/Fe] would
increase by the amount that the value of [Fe/H] is reduced.  However, one could
compensate for that by decreasing the assumed age.  For example, a 10 Gyr
isochrone for [Fe/H] $= -1.8$ and [$\alpha$/Fe] $= +0.02$ is essentially
indistinguishable from one for the same [Fe/H], but [$\alpha$/Fe] $= -0.13$;
see the right hand panel.)
 
This brings us to the second possibility, which is that stars of different
metal abundances have different ages (as suggested by our simulations in the
case of the oldest stars).  Fig.~12 shows that a 6 Gyr isochrone for
[Fe/H] $= -1.8$ and [$\alpha$/Fe] $= +0.4$ (middle panel) is bluer than a 4 Gyr
isochrone for [Fe/H] $= -1.4$ and [$\alpha$/Fe] $= +0.13$ (right-hand panel)
by only a small amount.  In fact, the latter would be nearly coincident with
the former if it had an age of 3 Gyr.  (This explanation would indeed imply
that ages vary with metallicity in just the right way to produce a thin giant
branch.)  In addition, variations in the heavy-element mixture may explain some
of the differences in the fits of isochrones to the $BV$ amd $VI$ photometry of
Carina giants.  It is an approximation to adopt the same abundances for all of
the $\alpha$-elements at a given [Fe/H] value.  As shown by V12, the location
of the RGB on the theoretical plane has a comparable dependence on the abundance
of Si as of Mg, and if, e.g., [Si/Fe] $<$ [Mg/Fe], then the values of
[$\alpha$/Fe] that we have assumed are too high, and the isochrones (especially
at higher metallicities) are too red.  Unfortunately, not a great deal is known
about the abundances of silicon in Carina.  (Star-to-star abundance variations
will also have important consequences for stellar temperatures and colors, but
color transformations that allow for differences in the abundances of single
elements are not yet available.)  At this stage, the uncertainties are still
large enough that the minor inconsistencies that exist between Figs.~11 and 12
are not of particular concern.  

\section{Summary}
\label{sec:sum}

The most important result of this investigation is that stellar models are
able to reproduce the main features of the CMD of Carina quite well when they
assume the full range in metal abundances ($\gta 1.5$ dex in [Fe/H]) that has
been derived from high-quality spectroscopic data.  This contradicts previous
findings, notably by Bono et al.~(2010), who argued from their analysis of the
observed pbotometry that the metallicity variations encompassed by either the
oldest or intermediate-age populations cannot amount to much more than a few
tenths of a dex.  The key to our success is that we have used isochrones (from
V14a) which assume heavy-element abundances that follow the observed relations
between [$\alpha$/Fe] and [Fe/H].  (Interpolations in those grids can be made
for any value of [$\alpha$/Fe] between $-0.4$ and $+0.4$ at a fixed [Fe/H]
value.)  In fact, it would not have been possible to reproduce some CMD
features, such as the small range in luminosity spanned by the old HB component
of Carina, using models for the measured range in [Fe/H] {\it unless}
[$\alpha$/Fe] is a strongly decreasing function of the iron abundance (as
observed).  In other words, models for constant [$\alpha$/Fe], irrespective of
the value that is adopted, are incapable of providing a satisfactory explanation
of the Carina CMD.
%(which is actually what Bono et al.~were telling us).

Being able to transform our models to the [$(B-V)_0,\,M_V$]- and
[$(V-I)_0,\,M_V$]-diagrams using fully consistent transformations (from CV14)
has certainly made a difference as well, because predicted $BVI$ magnitudes
have a considerable dependence on the abundances of the $\alpha$-elements at
fixed values of [Fe/H], $\log\,g$, and $\teff$.  Furthermore, we have taken the
additional step of determining, and applying, corrections to the CV14
$(B-V)$--$\teff$\ relations in order that isochrones reproduce the CMDs of M\,3
and M\,92, when very good estimates of their reddenings, metallicities, and
distances are assumed.  This ``calibration" of the isochrones ensures that
the models faithfully reproduce the locations of the MSTOs and lower RGBs, as 
well as the slopes of the SGBs in GCs that span a range of nearly 1 dex in
[Fe/H], thereby giving us confidence that they will accurately represent the
properties of stars in the dSph.  This procedure should minimize errors that
would otherwise be present, and possibly be quite substantial.  

The veracity of this assertion is easily verified.  There is a general
tendency of Victoria-Regina isochrones to suffer from small zero-point and
systematic offsets when compared with GC CMDs, though the predicted slopes of
the MS and RGB look fine (see, e.g., the plots provided by V13 and V14a).
When compared with the same {\it HST} observations that were the subject of the
paper by V13, Dartmouth isochrones appear to be considerably more problematic
(see Dotter et al.~2010; the right-hand panels of their Figs.~4 and 5).
Differences in the color transformations that were employed in the respective
studies are likely to be mostly responsible for this, in view of the fact that
very similar physics has been incorporated in both the Victoria and Dartmouth
codes.  However, the same cannot be said of BASTI isochrones (e.g., Pietrinferni
et al.~2004) because these computations do not take diffusive processes into
account, which have significant consequences for turnoff luminosity versus age
relations and the predicted $\teff$\ scale, especially at low metallicities
and high ages (see VandenBerg et al.~2002; their Fig.~2).  

It is risky to use stellar models from {\it any} source without carefully
examining how well they satisfy empirical constraints.  This is especially
true when attempting to interpret the photometry of a galaxy, like Carina.
When an isochrone is overlaid onto the observed CMD, it should superimpose
stars of nearly the same age and chemical abundances whether they are located
near the TO, along the SGB, or on the giant branch.  This is, of course, 
complicated by known degeneracies between the effects of age and chemical
abundances, but if the adopted isochrones do not reproduce the locations and
morphologies of GC CMDs particularly well, those models are unlikely to
provide very meaningful results when applied to galactic observations.  (GC
ages, in contrast, mainly involve fits of isochrones to turnoff stars.)

The ages of the different stellar populations in Carina that we obtained
from overlays of isochrones onto the observed CMD are quite similar to those
found previously, especially in a relative sense.  The oldest of the
intermediate-age stars appear to have formed 6--6.5 Gyr ago, whereas the range
in age needed to reconcile stellar models with the faintest turnoff/SGB stars
is $\sim 9$--14 Gyr.  In agreement with earlier determinations, there was
apparently a $\sim 3$--4 Gyr hiatus in the formation of stars in Carina after
the first SF epoch.

Our CMD simulations, which are based on isochrones that have been carefully
calibrated, rule out the possibility that the faintest TO/SGB stars in Carina
are a single-age population by more than $6\,\sigma$.  However, simulations
in which ages are limited to the range 8--14.4 Gyr and iron abundances to
$-3.0 \le$ [Fe/H] $\le -1.4$, without constraining how the number of stars
varies with metallicity, provide a significantly better match to the observed
CMD if the earliest star formation epoch lasted a few Gyr.  Moreover, judging
from the derived weighting of the points in a grid of age and [Fe/H], there are
two ways of obtaining a reasonable approximation to the observed MDF (both of
which suggest that ages decrease somewhat with increasing [Fe/H]).  Either
the adopted distance modulus, $(m-M)_V = 20.28$, should be increased by $\sim
0.1$ mag, if [$\alpha$/Fe] varies with [Fe/H] according to the relation shown
in Fig.~1a, or the assumed O abundances should be increased and the dependence
of [O/H] on [Fe/H] is similar to that adopted by Brown et al.~(2014).  In
support of the latter possibility, we note that [O/Fe] $\gta +0.6$ is typically
found in very metal poor, Galactic halo stars (e.g., Garc\'ia P\'erez et
al.~2006, Ram\'irez et al.~2012), and as discussed in \S~\ref{subsec:TO}, it
may be difficult to reconcile the distribution of stars along the old HB
component of Carina with the observed MDF unless the stars have relatively
high O abundances.  (A larger distance seems unlikely in view of the recent
empirical determinations by Pietrzy\'nski et al.~2009 and Coppola et al.~2013.)

Although our isochrones had no difficulty reproducing the narrow RGB of Carina
on the $V-I,\,V$ CMD, the observed width on the $B-V,\,V$ color plane was
found to be smaller than expected.  Improved consistency would be obtained if
the age-metallicity relations that describe the stellar populations in Carina
have significant slopes, in the expected sense that more metal-rich stars are
younger.  In fact, this is suggested by the simulations that we, and de Boer
et al.~(2014b), have carried out for the oldest, and intermediate-age,
populations of Carina, respectively.  (Our simulations were limited to the
oldest stars because extensions of the V14a grids to younger ages are not
available.)  In addition, or alternatively, this difficulty may be a reflection
of current uncertainties associated with the observed [Fe/H] scale, the assumed
[$\alpha$/Fe] values (which are based entirely on Mg, as little is known about
the abundances of Si), and/or the color transformations.  While a resolution of
this problem must be left for future work, it is very encouraging that modern
stellar models, on the assumption of the chemical abundances which have been
derived from high-resolution spectroscopy, are able to explain most of the
features of Carina's CMD quite well.

\acknowledgements

We thank both Chris Pritchet, who pointed out a very useful reference, and Kim
Venn for helpful discussions.  D.A.V acknowledges the support of a Discovery
Grant from the Natural Sciences and Engineering Research Council of Canada.


\begin{references}
%
%\reference{}
%Asplund, M., Grevesse, N., Sauval, A.~J., \& Scott, P.~2009, ARAA, 47, 481
%
%\reference{}
%Asplund, M., Lambert, D.~L., Nissen, P.~E., Primas, F., \& Smith, V.~V.~2006,
% ApJ, 644, 229

\reference{}
Battaglia, G., Irwin, M.~J., Tolstoy, E., de Boer, T., \& Mateo, M.~2012,
 ApJ, 761, L31
%
%\reference{}
%Bennett, C.~L., Larson, D., Weiland, J.~L., et al.~2013, ApJS, 208, 20

\reference{}
Bono, G., Stetson, P.~B., Walker, A.~R., et al.~2010, PASP, 122, 651

\reference{}
Brown, T.~M., Smith, E., Ferguson, H.~C., et al.~2006, ApJ, 652, 323

\reference{}
Brown, T.~M., Tumlinson, J., Geha, M., et al.~2012, ApJ, 735, L21

\reference{}
Brown, T.~M., Tumlinson, J., Geha, M., et al.~2014, ApJ, 796, 91

\reference{}
Carretta, E., Bragaglia, A., Gratton, R.~G., D'Orazi, V., \& Lucatello,
 S.~2009a, A\&A, 508, 695

\reference{}
Carretta, E., Bragaglia, A., Gratton, R.~G., \& Lucatello, S.~2009b, A\&A,
 505, 139

\reference{}
Carretta, E., \& Gratton, R.~G.~1997, A\&AS, 121, 95

\reference{}
Casagrande, L., \& VandenBerg, D.~A.~2014, MNRAS, 444, 392\ \ \ (CV14)

\reference{}
Castellani, V., \& Degl'Innocenti, S.~1995, A\&A, 298, 827

\reference{}
Cayrel, R., Depagne, E., Spite, M., et al.~2004, A\&A, 416, 1117

\reference{}
Coppola, G., Stetson, P.~B., Marconi, M., et al.~2013, ApJ, 775, 6

\reference{}
de Boer, T.~J.~L., Belokurov, V., Beers, T.~C., \& Lee, Y.~S.~2014a, MNRAS,
 443, 658

\reference{}
de Boer, T.~J.~L., Tolstoy, E., Lemasle, B., Saha, A., Olszewski, E.~W.,
 Mateo, M., Irwin, M.~J., \& Battaglia, G.~2014b, A\&A, 572, A10

\reference{}
Dolphin, A.~E.~2002, MNRAS, 332, 91
%
%\reference{}
%Dorman, B., \& Rood, R.~T.~1993, ApJ, 409, 387

\reference{}
Dotter, A., Sarajedini, A., Anderson, J., et al.~2010, ApJ, 708, 698

\reference{}
Fabbian, D., Nissen, P.~E., Asplund, M., Pettini, M., \& Akerman, C.~2009,
 A\&A, 500, 1143

\reference{}
Fabrizio, M., Merle, T., Th\'evenin, F., et al.~2012, PASP, 124, 519

\reference{}
Frebel, A.~2010, AN, 331, 474

\reference{}
Garc\'ia P\'erez, A.~E., Asplund, M., Primas, F., Nissen, P.E., \& Gustafsson,
 B.~2006, A\&A 451, 621

\reference{}
Gilmore, G.~F., \& Wyse, R.~F.~G.~1991, ApJ, 367, L55

\reference{}
Gilmore, G.~F., \& Wyse, R.~F.~G.~1998, AJ, 116, 748
%
%\reference{}
%Gustafsson, B., Edvardsson, B., Eriksson, K., Jorgensen, U.~G., Nordlund, 
%\AA, \& Plez, B.~2008, A\&A, 486, 951

\reference{}
Harris, J., \& Zaritsky, D.~2001, ApJS, 136, 25

\reference{}
Hurley-Keller, D., Mateo, M., \& Nemec, J.~1998, AJ, 115, 1840

\reference{}
Kirby, E.~N., Cohen, J.~G., Smith, G.~H., Majewski, S.~R., Sohn, S.~T., \&
 Guhathakurta, P.~2011, ApJ, 727, 79

\reference{}
Koch, A., Grebel, E.~K., Wyse, R.~F.~G., et al.~2006, AJ, 131, 895

\reference{}
Koch, A., Grebel, E.~K., Gilmore, G.~F., et al.~2008, AJ, 135, 1580

\reference{}
Komatsu, E., Smith, K.~M., Dunkley, J., et al.~2011, ApJS, 192, 18

\reference{}
Kraft, R.~P., \& Ivans, I.~I.~2003, PASP, 115, 143

\reference{}
Lamb, M.~P., Venn, K.~A., Shetrone, M.~D., Sakari, C.~M., \& Pritzl, 
 B.~J.~2015, MNRAS, 448, 42

\reference{}
Landolt, A.~1973, AJ, 78, 959

\reference{}
Landolt, A.~1992, AJ, 104, 340

\reference{}
Lemasle, B., Hill, V., Tolstoy, E., et al.~2012, A\&A, 538, A100

\reference{}
Lianou, S., Grebel, E.~K., \& Koch, A.~2011, A\&A, 531, A152

\reference{}
Maoz, D., Mannucci, F., \& Nelemans, G.~2014, ARA\&A, 52, 107

\reference{}
McWilliam, A.~1997, ARA\&A, 35, 503
%
%\reference{}
%Mighell, K.~J.~1990, A\&AS, 82, 1

\reference{}
Minor, Q.~E.~2013, ApJ, 779, 116

\reference{}
Monelli, M., Milone, A.~P., Fabrizio, M., et al.~2014, ApJ, 796, 90

\reference{}
Pietrinferni, A., Cassisi, S., Salaris, M., \& Castelli F.~2004, ApJ, 612, 168

\reference{}
Pietrzy\'nski, G., G\'orski, M., Gieren, W., Ivanov, V.~D.,  Bresolin, F., \&
Kudritzki, R.-P.~2009, AJ, 138, 459

\reference{}
Ram\'irez, I., Mel\'endez, J., \& Chanam\'e, J.~2012, ApJ, 757, 164
%
%\reference{}
%Ram\'irez, I., Allende, Prieto, C., \& Lambert, D.~L.~2013, ApJ, 764, 78

\reference{}
Roederer, I.~U., \& Sneden, C.~2011, AJ, 142, 22

\reference{}
Sarajedini, A., Bedin, L.~R., Chaboyer, B., et al.~2007, AJ 133, 1658

\reference{}
Schlafly, E.~F., \& Finkbeiner, D.~P.~2011, ApJ, 737, 103

\reference{}
Shetrone, M.~D., Venn, K.~A., Tolstoy, E., Primas, F., Hill, V., \& Kaufer,
 A.~2003, AJ, 125, 684

\reference{}
Smecker-Hane, T.~A., Stetson, P.~B., Hesser, J.~E., \& Lehnert, M.~D.~1994,
 AJ, 108, 507

\reference{}
Sobeck, J.~A., Kraft, R.~P., Sneden, C., et al.~2011, AJ, 141, 175

\reference{}
Starkenburg, E., Hill, V., Tolstoy, E., et al.~2010, A\&A, 513, A34

\reference{}
Stetson, P.~B.~2000, PASP, 112, 925  

\reference{}
Stetson, P.~B.~2005, PASP, 117, 563

\reference{}
VandenBerg, D.~A., \& Bell, R.~A.~2001, New Astr.~Rev., 45, 577

\reference{}
VandenBerg, D.~A., Bergbusch, P.~A., Dotter, A., Ferguson, J.~W., Michaud, G.,
Richer, J., \& Proffitt, C.~R.~2012, ApJ, 755, 15\ \ \ (V12)

\reference{}
VandenBerg, D.~A., Bergbusch, P.~A., Ferguson, J.~W., \& Edvardsson, B.~2014a, 
ApJ, 794, 72\ \ \ (V14a)

\reference{}
VandenBerg, D.~A., Bond, H.~E., Nelan, E.~P., Nissen, P.~E., Schaefer, G.~H.,
\& Harmer, D.~2014b, ApJ, 792, 110\ \ \ (V14b)

\reference{}
VandenBerg, D.~A., Brogaard, K., Leaman, R., \& Casagrande, L.~2013, ApJ, 775,
 134\ \ \ (V13)

\reference{}
VandenBerg, D.~A., \& Clem, J.~L.~2003, AJ, 126, 778

\reference{}
VandenBerg, D.~A., Richard, O., Michaud, G., \& Richer, J.~2002, ApJ, 571, 487

\reference{}
VandenBerg, D.~A., Swenson, F.~J., Rogers, F.~J., Iglesias, C.~A., \&
 Alexander, D.~R.~2000, ApJ, 532, 430

\reference{}
Venn, K.~A., Irwin, M., Shetrone, M.~D., Tout, C.~A., Hill, V., \& Tolstoy,
 E.~2004, AJ, 128, 1177

\reference{}
Venn, K.~A., Shetrone, M.~D., Irwin, M.~J., et al.~2012, ApJ, 751, 102

\reference{}
Vivas, A.~K., \& Mateo, M.~2013, AJ, 146, 141

\end{references}
\end{document}